\documentclass[aps,prl,longbibliography,twocolumn,superscriptaddress]{revtex4-2}
\usepackage{amsmath,amssymb,bm,graphicx,color,bbold,hyperref,keyval,url,latexsym,dsfont}
\usepackage{xcolor}
\usepackage{booktabs}
\usepackage[normalem]{ulem} 
\usepackage{CJK}
\usepackage{float}
\usepackage{comment}

\definecolor{chg}{HTML}{2E7D32}

\begin{document}
\title{Competing states in the $S=1/2$ triangular-lattice $J_1$-$J_2$ Heisenberg model: a dynamical density-matrix renormalization group study}

\begin{CJK*}{UTF8}{}
\author{Shengtao Jiang (\CJKfamily{gbsn}蒋晟韬)}
\altaffiliation{contact author: stjiang@stanford.edu}
\affiliation{Stanford Institute for Materials and Energy Sciences, SLAC National Accelerator Laboratory and Stanford University, Menlo Park, CA 94025, USA}

\author{Steven R. White}
\affiliation{Department of Physics and Astronomy, University of California, Irvine, California 92697, USA}

\author{Steven A. Kivelson}
\affiliation{Department of Physics, Stanford University, Stanford, CA 94305, USA}

\author{Hong-Chen Jiang}
\altaffiliation{contact author: hcjiang@stanford.edu}
\affiliation{Stanford Institute for Materials and Energy Sciences, SLAC National Accelerator Laboratory and Stanford University, Menlo Park, CA 94025, USA}

\date{\today}

\begin{abstract}
Previous studies of the $S=1/2$ triangular-lattice $J_1$--$J_2$ Heisenberg antiferromagnet have inferred the existence of a non-magnetic ground-state phase for an intermediate range of $J_2$, but disagree concerning whether it is a gapped $\mathbb{Z}_2$ quantum spin liquid (QSL), a gapless (Dirac) QSL, or a weakly symmetry-broken phase. Using an improved dynamical density-matrix renormalization group method, we investigate the relevant intermediate $J_2$ regime for cylinders with circumferences from 6 to 9. Depending on the initial state and boundary conditions, we find two {\it distinct} variational states. The higher energy state is consistent with a Dirac QSL. In the lower-energy state, both the static and dynamical properties are qualitatively similar to the magnetically ordered state at $J_2=0$,  suggestive of either a weakly magnetically ordered non-QSL or a gapped QSL proximate to a continuous transition to such an ordered state. 
\end{abstract}
\maketitle
\end{CJK*}

{\it Introduction.}--
Half a century ago, Anderson initiated the study of quantum spin liquids (QSL) with his seminal proposal of the resonating valence bond state in the triangular-lattice antiferromagnet (AFM)~\cite{rvb,shivajiandme}, prototypical of what is now known as a gapped $\mathbb{Z}_2$ QSL~\cite{Broholm_2020}.
In this exotic state of matter, strong quantum fluctuations, facilitated by geometrical frustration in the triangular lattice, preclude conventional magnetic order even at zero temperature. Despite the later discovery that the ground state of the nearest-neighbor (NN) Heisenberg AFM has the classical 120$^\circ$ N\'eel order~\cite{tri-neel1,tri-neel2,tri-neel3,tri-neel4,tri-neel5,steve-sasha-07}, the triangular lattice AFM remains one of the central focuses in the search for QSL and other exotic quantum states, both theoretically~\cite{laughlin,sondhichandra,tri-cheng,tri-donna,tri-hub-dmrg,tri-sasha1,tri-sasha2,tri-sasha3,tri-xiang,j1j2-cesar,tri-mingpu,tri-tohyama,tri-liwei} and experimentally~\cite{tri-exp1,tri-exp2,tri-exp3,tri-exp4,tri-exp5,tri-exp6,tri-exp7,tri-exp8,tri-exp9,tri-exp10,tri-exp11,tri-exp12,tri-exp13,tri-exp14,tri-exp15}.

\begin{figure*}[t]
  \includegraphics[width= 0.8\linewidth]{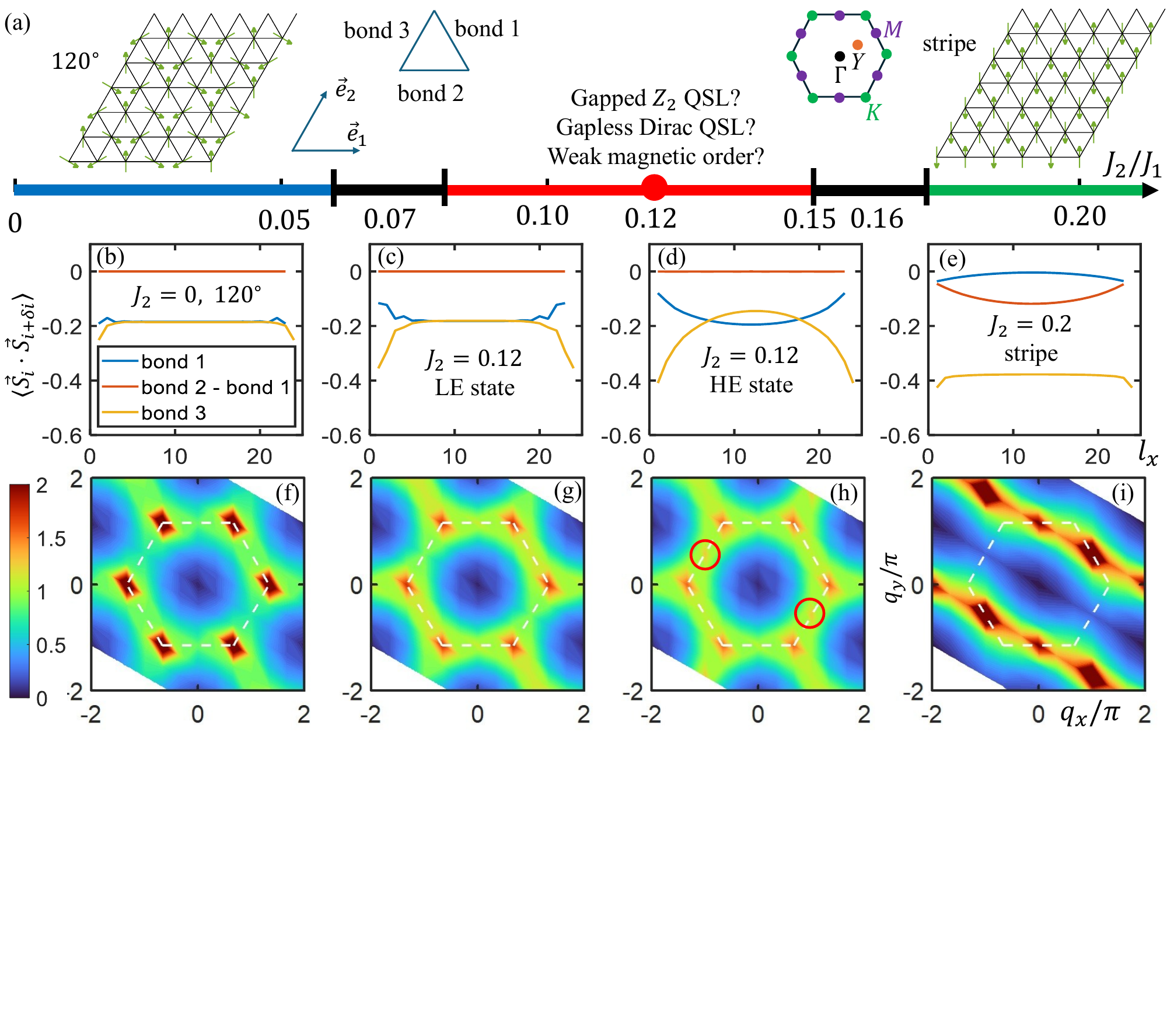}
  \vskip -4.0cm
  \caption{
  {\bf Top panel} (a): Conjectured ground state phase diagram of the $S=1/2$ $J_1$-$J_2$ Heisenberg model on the triangular lattice from previous studies~\cite{j1j2-zhu,j1j2-becca1,j1j2-cenke,j1j2-mcculloch,j1j2-donna,j1j2-campbell,j1j2-donna2,j1j2-tom,j1j2-lauchli,j1j2-imada,j1j2-hc}, where an intermediate $0.07 \lesssim J_2\lesssim 0.16$ region was found with vanishing magnetic order, albeit with somewhat different phase boundaries and different putative QSLs.
  The middle point of this region $J_2=0.12$, and specifically what QSL if any occurs there, is the focus of this paper. 
  Insets of (a) show the XC6 cylinder 
  with $\vec e_1$ and $\vec e_2$ being the unit vectors along its length and circumference, respectively, and the first Brillouin Zone with the high symmetry points marked. The $Y$ point (midpoint between $\Gamma$ and $M$) is where the Dirac spinon node is located for a gapless Dirac QSL~\cite{dirac-theory,dirac-theory2}.
  {\bf Middle panel}: Strength of the spin-spin correlation $\langle \vec S_i\cdot \vec S_{i+\delta i}\rangle$ on three types of NN bonds (defined in (a)), averaged over the circumferential direction and plotted versus the length of the cylinder $l_x$, for four different states: (b) the 120$^\circ$ state at $J_2=0$, (c) the LE state at $J_2=0.12$, (d) the HE state at $J_2=0.12$ and (e) the striped state at $J_2=0.2$. The strengths of bonds 1 and 2 are identical except in (e). 
  {\bf Bottom panel}: The equal-time spin structural factor $S(\vec q)$ for (f) the 120$^\circ$ state at $J_2=0$ that has dominant peaks at $K$, (g) the LE state at $J_2=0.12$ with weaker peaks at $K$, (h) the HE state at $J_2=0.12$ with still weaker peaks at both $K$ and $M$ (marked by red circles), and (i) the striped state at $J_2=0.2$ with peaks at $M$. The white dashed lines mark the boundaries of the first Brillouin zone.}
  \label{fig:phd}
\end{figure*}

While the simplest NN Heisenberg model fails to achieve a QSL state, quantum fluctuations are apparently enhanced by the inclusion of an AFM second NN exchange $J_2$. Numerous studies~\cite{j1j2-zhu,j1j2-becca1,j1j2-cenke,j1j2-mcculloch,j1j2-donna,j1j2-campbell,j1j2-donna2,j1j2-tom,j1j2-lauchli,j1j2-imada,j1j2-hc,j1j2-oleg,j1j2-pollmann3} report a nonmagnetic phase, frequently interpreted as a QSL, in an intermediate $J_2$ range $0.07\lesssim J_2\lesssim0.16$, although the precise phase boundaries and whether the putative QSL is gapped or gapless remain unsettled.
Several density-matrix renormalization group (DMRG) studies~\cite{j1j2-zhu,j1j2-donna,j1j2-mcculloch,j1j2-hc} produced suggestive evidence of a gapped QSL, leading to the proposed phase diagram sketched in Fig.~\ref{fig:phd}(a).
Specifically, that the QSL has an emergent $\mathbb{Z}_2$ gauge symmetry was inferred from the existence of two low-energy states on even-width cylinders, depending on boundary conditions  
-- a feature that was interpreted as representing the two-fold topological degeneracy\cite{thouless} expected of a ``$\mathbb{Z}_2$ QSL'' on a cylinder.
Additionally, a substantial spin triplet gap on finite-width cylinders was observed.
 
However, on the basis of DMRG simulations
with flux insertion~\cite{j1j2-yche} and variational Monte Carlo~\cite{j1j2-becca1,j1j2-imada}, some subsequent studies argued that the relevant QSL is a gapless ``Dirac QSL.''
This view was further supported by studies that 
found evidence of low-energy magnetic excitations at the $M$ points in the Brillouin zone (illustrated in Fig.~\ref{fig:phd}(a)) - consistent with the existence of nodal spinons at the $Y$ points~\cite{j1j2-becca-prx,j1j2-moore,j1j2-pollmann,j1j2-pollmann-2025} - and by other more complicated evidence of emergent $U(1)$ gauge fields~\cite{qed3-lauchli}. 
On the other hand, the latest DMRG simulations find that the spin triplet gap remains substantial even for cylinders with widths up to 10~\cite{j1j2-hc}, behavior that is also seen in the spectral function on a six-leg cylinder~\cite{j1j2-moore,j1j2-pollmann}.

Thus, several fundamental questions remain concerning the nature of the intermediate $J_2$ region. 
1) Are the two nearly degenerate states seen in previous studies a reflection of the expected topological degeneracies of a $\mathbb{Z}_2$ QSL or do they reflect the existence of two nearly degenerate, thermodynamically distinct ``competing'' ground-state phases?  
2) Correspondingly, how do the two nearly degenerate states and their associated spectral properties evolve as the system size increases?
3) Most importantly, is the intermediate regime a QSL in the two-dimensional limit, or does it ultimately develop weak magnetic or valence-bond order?

In this Letter, we address these questions by systematically examining both the ground-state and dynamical properties of the $S=1/2$ triangular-lattice $J_1$-$J_2$ Heisenberg model in the intermediate $J_2$ regime using large-scale DMRG~\cite{White1992,White1993} and dynamical DMRG~\cite{ddmrg-jeckelmann} simulations, focusing on the representative value $J_2 = 0.12$ within the intermediate regime.  
Among other things, we establish the existence of two low-energy competing states that can be obtained from different initial conditions. This behavior is observed on cylinders with circumferences ranging from 6 to 9, i.e., it is not restricted to even-width cylinders as would be expected for a $\mathbb{Z}_2$ QSL. 
To obtain dynamical information about the various quantum phases, we have improved and then utilized the dynamical DMRG~\cite{ddmrg-jeckelmann} to probe their dynamical spin structure factors (DSSF).
In this way, we have uncovered qualitatively distinct low-energy excitation spectra for the two states.
Thus, in answer to question 1, the existence of two nearly degenerate states does {\it not} reflect a topological signature of a gapped $\mathbb{Z}_2$ QSL. 
Instead, we will show that the higher energy state (HE state) most likely represents a Dirac QSL, characterized by low-frequency peaks of comparable intensity at both the $K$ points and $M$ points in the DSSF.
In contrast, the lower energy state (LE state) exhibits both static and dynamical structure factors characterized by a single dominant peak at the $K$ points, indicating it is probably a gapped $\mathbb{Z}_2$ spin liquid, but could be a weakly ordered version of the $J_2=0$ phase close to a quantum critical point.

Thus, while overall our findings lend support to the existence of a gapped $\mathbb{Z}_2$ QSL as sketched in Fig.~\ref{fig:phd}(a), the possibility that the character of the ground-state may change for still larger $L_y$ is non-negligible, i.e., questions 2 and 3 can only be answered very tentatively.  Specifically, while the Dirac QSL is the higher energy state (and hence at best metastable), this could reverse with larger $L_y$~\cite{vmc-cyliner}, as previous studies show that the energy density difference is only of order 0.1\% and shrinks as $L_y$ increases~\cite{j1j2-donna,j1j2-zhu}.  Moreover, even accepting that the low energy state is best characterized as a $\mathbb{Z}_2$ QSL, because it appears close to a magnetically ordered state, its character could easily change with larger $L_y$.

{\it Model and ground state properties.}--
We study the $S=1/2$, $J_1$-$J_2$ Heisenberg model on triangular cylinders:
\begin{equation}
    \label{eq:j1j2mod}
    H=J_1\sum_{\langle ij \rangle_1} \vec{S}_i\cdot \vec {S}_j + J_2\sum_{\langle ij \rangle_2} \vec{S}_i\cdot \vec {S}_j,
\end{equation}
where $\langle ij \rangle_{1/2}$ denotes the first/second NN pairs of sites, and $\vec S_j$ are spin-1/2 operators.
$J_1=1.0$ is set as the energy unit throughout the paper.
We primarily use the so-called XC cylinders, depicted in Fig.~\ref{fig:phd}(a), where one of the NN bonds is along the length of the cylinder with unit vector $\vec{e}_1=(1,0)$. The circumference of the cylinder is along $\vec{e}_2=(\frac{1}{2},\frac{\sqrt3}{2})$, meaning that sites are connected periodically along this direction. We also consider YC cylinders for comparison and find similar results~\cite{sm}.

One of the major findings of previous DMRG studies was the existence, in the intermediate $J_2$ region, of two distinct states on even-width cylinders~\cite{j1j2-zhu,j1j2-donna}. 
These were often referred to as the ``odd'' and ``even'' states, since it was conjectured that they represent the ground states in the two topological sectors expected in a $\mathbb{Z}_2$ QSL on a cylinder~\cite{z2-tri-moessner,z2-theory-sachdev,z2-theory-ashvin, thouless}. 
Since our findings do not support this picture, we will refer to them as the LE and HE states instead. 

For a given system, the two states are found by starting DMRG with different initial states. 
For example, the LE state can always be reached starting with the $J_2=0$ ground state, while the HE state is accessed starting from a striped product state. 
The LE state can also be stabilized by isolating one site on each boundary of the cylinder, which might be thought of as corresponding to localizing a spinon at each end, thus favoring $\mathbb{Z}_2$ QSL relative to any confined phase.
The energy density difference between the two states is of order $10^{-3}$, and shrinks as $L_y$ increases~\cite{j1j2-zhu,j1j2-donna}. 
The HE state is metastable in the sense that further sweeps at the bond dimension accessed or listed do not result in its conversion into the LE state~\footnote{We have also tested that the HE state on XC6 cylinder remains metastable upon adding a noise of 1E-5 in DMRG.}. 

The existence of two topologically distinct ground states with energies that approach each other exponentially is an expected feature of a gapped $\mathbb{Z}_2$ QSL.
However, the big differences in the locally observable properties of the HE and LE states 
are inconsistent with this characterization.
As shown in Figs.~\ref{fig:phd}(b) - (e), the spin-spin correlations on the NN bonds with the three possibly orientations are similar in the bulk for both the LE state and the 120$^\circ$ state, while for the HE state and the striped state, one orientation of the bonds encircling the cylinder is noticeably different from the others, i.e. there is an indication of vestigial $Z_3$ nematic order that might reflect proximity to a striped state.
Moreover, comparing the equal-time spin structure factors $S(\vec q)$ in Fig.~\ref{fig:phd}(g) and (h), we see that the HE state has extra peaks at two of the $M$ points $\pm (\pi, \pi/\sqrt3)$, which are not seen in the LE state.
If we ignore the amplitude and focus on the sign pattern of the longer-ranged correlations, we find a more drastic contrast, suggesting that their low-energy excitations are essentially different, see the End Matter.
Additionally, in contrast with previous studies~\cite{j1j2-zhu} and the analysis~\cite{yao,j1j2-zhu} of the quantum dimer model of a $\mathbb{Z}_2$ QSL, we observe the existence of a metastable distinct HE state also on odd-width XC9 (see the End Matter) and XC7~\cite{sm} cylinders.

While our results clearly show that the LE and HE states are different in 
qualitative, locally observable ways that preclude a topological interpretation, it is notable that the LE state shares many similarities with the 120$^\circ$ state at $J_2=0$: they both have pronounced peaks in their structure factors at the $K$ points in the Brillouin zone, they both support edge spinons on even-width cylinders, weak columnar dimer order on odd-width cylinders (albeit significantly weaker in the $J_2=0$ state), as well as a similar $L_y$ dependence of the spin triplet gap~\cite{sm}.  Both the differences between the LE and HE states, and the similarities between the LE state and the 120$^\circ$ state at $J_2=0$ are more clearly reflected in the dynamical correlations to which we now turn.

{\it Improved Dynamical DMRG method.}--
Having investigated the ground state properties of the LE and the HE states, we aim to probe their spectral properties through the DSSF $S(\vec q,\omega)$, defined as:
\begin{equation}
\begin{split}
S(\vec q,\omega)=&\frac{-1}{\pi N}\sum_{\substack{a,b=1\\\alpha=x,y,z}}^{N}e^{i\vec{q}\cdot(\vec{r}_a-\vec{r}_b)}\times\\
&\text{Im}\langle 0|S^\alpha_{\vec{r}_a}\frac{1}{E_0+\omega-H+i\eta}S^\alpha_{\vec{r}_b}|0\rangle,
\label{eq:dssf}
\end{split}
\end{equation}
where $|0\rangle$ is the ground state obtained by DMRG with energy $E_0$, and $\eta$ is the chosen broadening factor. 
Dynamical-DMRG (DDMRG)~\cite{ddmrg-jeckelmann} computes this quantity by finding the complex correction vector $|c^\alpha_{\vec{r}_b}(\omega)\rangle=|X^\alpha_{\vec{r}_b}(\omega)\rangle+i|Y^\alpha_{\vec{r}_b}(\omega)\rangle$ that satisfies:
\begin{equation}
(E_0+\omega-H+i\eta)|c^\alpha_{\vec{r}_b}(\omega)\rangle=S^\alpha_{\vec{r}_b}|0\rangle.
\label{eq:corr_vec}
\end{equation}
In this way, DDMRG computes the DSSF in the frequency domain directly, with each calculation targeting one specific frequency. Typically, the low-frequency spectra are easier to obtain since the corresponding correction vector is closer to the ground state and has lower entanglement. 
In fact, it has been shown that DDMRG can be more efficient than time-dependent DMRG in obtaining spectra at a specific frequency~\cite{ddmrg++}.

Unlike DMRG, DDMRG, when implementing $H^2$ as two applications of $H$, is non-variational and solves a projected correction-vector equation, whose accuracy depends on the completeness of the truncated matrix product state (MPS) basis. 
Specifically, the standard state-averaged implementation~\cite{ddmrg-jeckelmann} must accommodate four distinct targets---$|0\rangle$, $S^\alpha_{\vec{r}_b}|0\rangle$, $|X^\alpha_{\vec{r}_b}(\omega)\rangle$, and $|Y^\alpha_{\vec{r}_b}(\omega)\rangle$---within a single truncated MPS basis; with limited bond dimension $m$, the resulting truncation can significantly distort the linear problem, leading to slow or spurious convergence. See the End Matter for a more detailed discussion.

After our own experimentation with different implementations, we found the following reliable and efficient approach to mitigate this issue, where we (i) decrease the number of simultaneously targeted states in state-averaging and (ii) re-balance their contributions during state averaging. Following Ref.~\cite{ddmrg++}, we keep $|0\rangle$ and the source state $S^\alpha_{\vec{r}_b}|0\rangle$ as a separate (fixed) MPS, and remove $|0\rangle$ from the state-averaging used to build the correction-vector basis. We then construct the state-averaged density matrix from three targets, $\{\sqrt{\eta}\,S^\alpha_{\vec{r}_b}|0\rangle,\ \eta|X^\alpha_{\vec{r}_b}(\omega)\rangle,\ |Y^\alpha_{\vec{r}_b}(\omega)\rangle\}$, where the $\eta$-dependent prefactors set their relative weights in the truncation. 
Empirically, we found this balancing stabilizes convergence and enables correct qualitative features of the DSSF to be captured at a moderate bond dimension, typically within twice of that for the ground state~\footnote{For ground state DMRG, we typically keep a maximum bond dimension such that the truncation error is $~\sim O(10^{-5})$.}, even though the truncation error is still $\sim O(10^{-4})$
We typically choose a tolerance of $\eta$ and a maximum iteration of 30 in the conjugate gradient optimization at each step, and perform DDMRG sweeps until convergence.  
More detailed information regarding the original and improved DDMRG method, including its benchmark and convergence, is provided in the SM~\cite{sm}. We implemented the improved DDMRG algorithm using the ITensor library~\cite{itensor} and Krylovkit~\cite{krylovkit}.

\begin{figure}[t]
	\includegraphics[width= 1.0\linewidth]{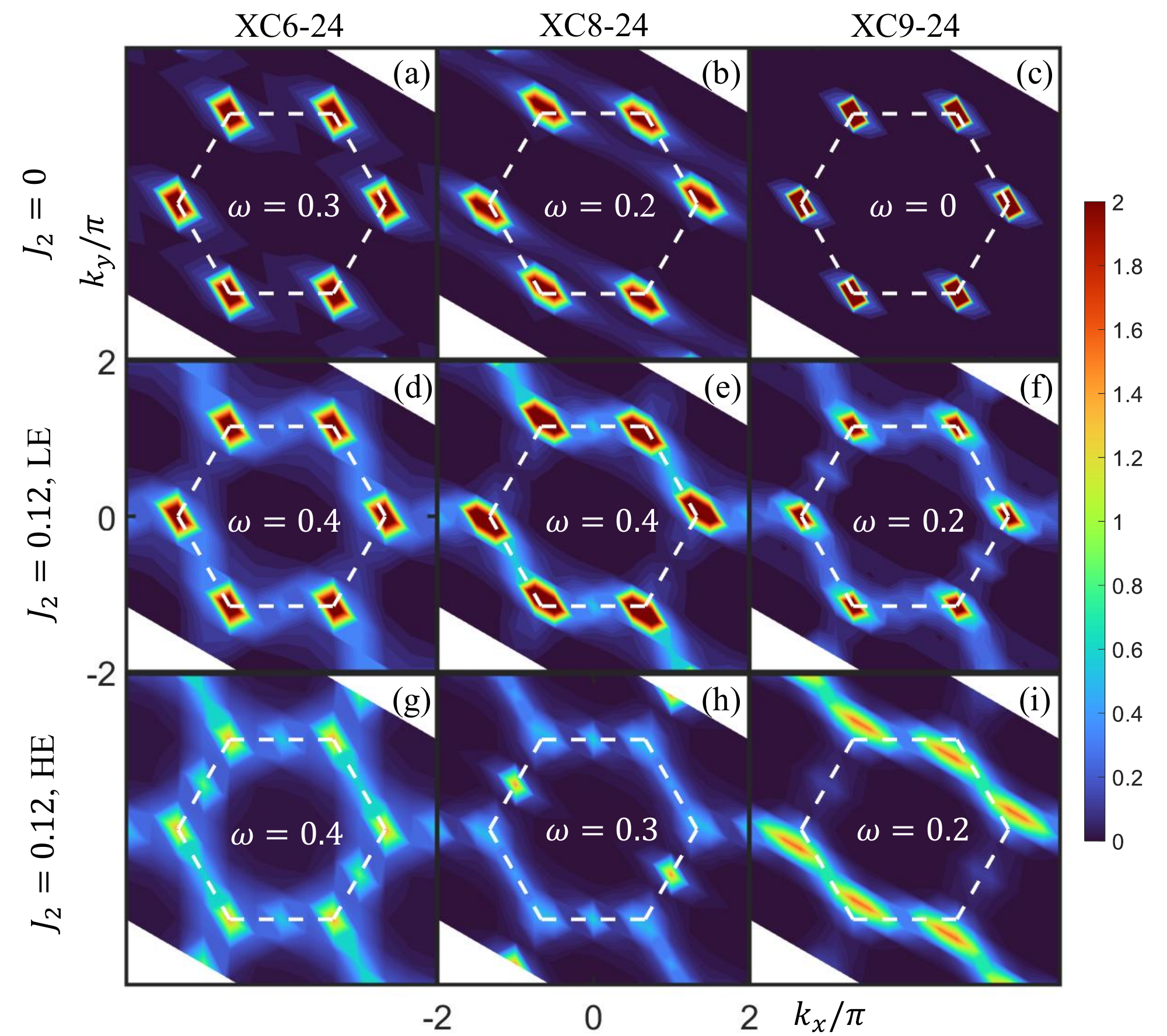}
	\caption{The low-energy DSSF $S(\vec q,\omega) $ at a specific $\omega$ shown in the center of each figure, for the 120$^\circ$ state at $J_2=0$ (top row, (a)(b)(c)), the LE state at $J_2=0.12$ (middle row, (d)(e)(f)), and the HE state at $J_2=0.12$ (bottom row, (g)(h)(i)). Simulations were carried out on XC6-24 (left column, (a)(d)(g)), XC8-24 (middle column, (b)(e)(h)), and XC9-24 (right column, (c)(f)(i)), respectively. The color scale has an upper cutoff of 2.}
	\label{fig:sqw}
\end{figure}

\begin{figure}[ht]
	\includegraphics[width=1.0\linewidth]{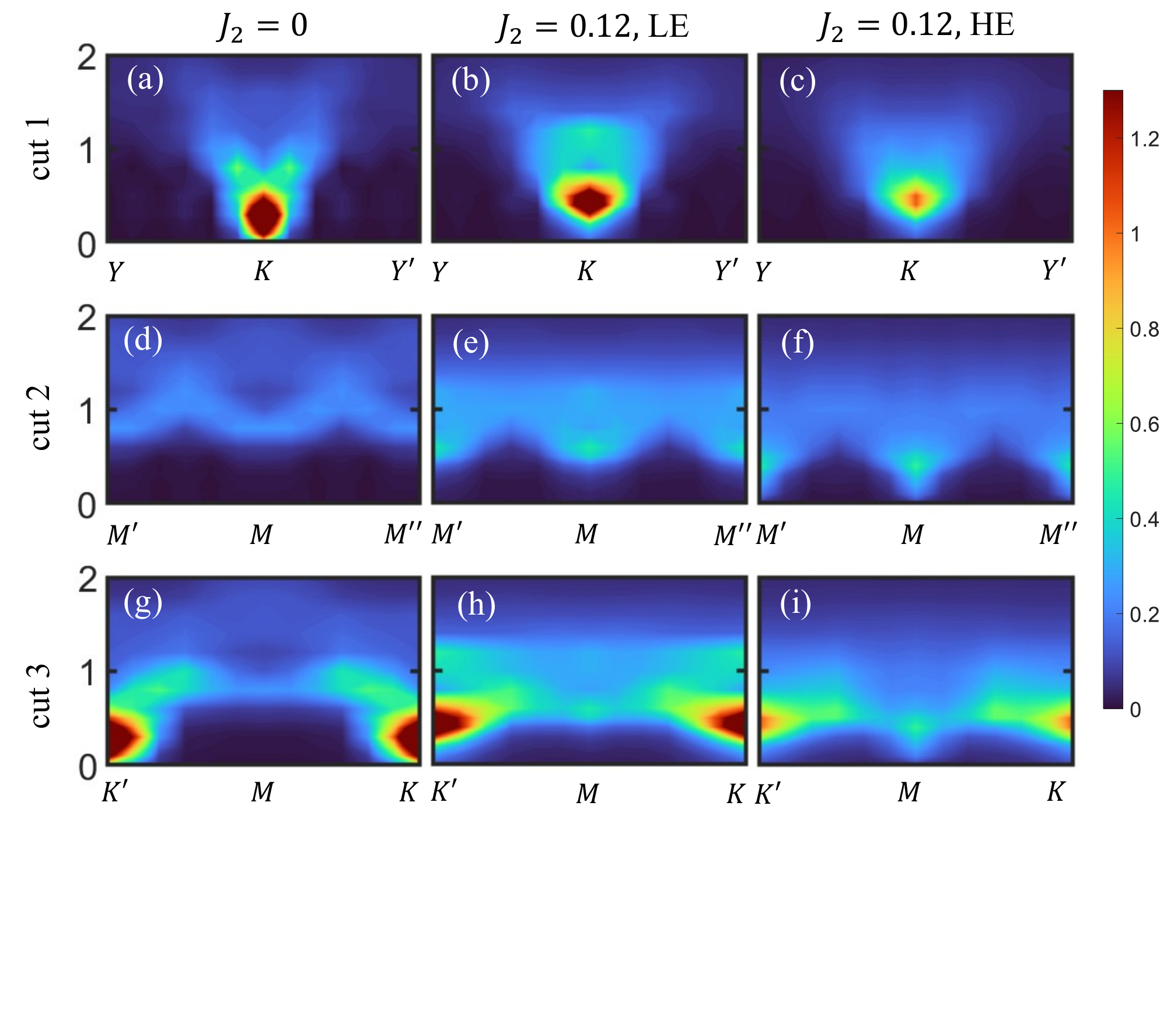}
    \vskip -1.5cm
    \includegraphics[width=1.0\linewidth]{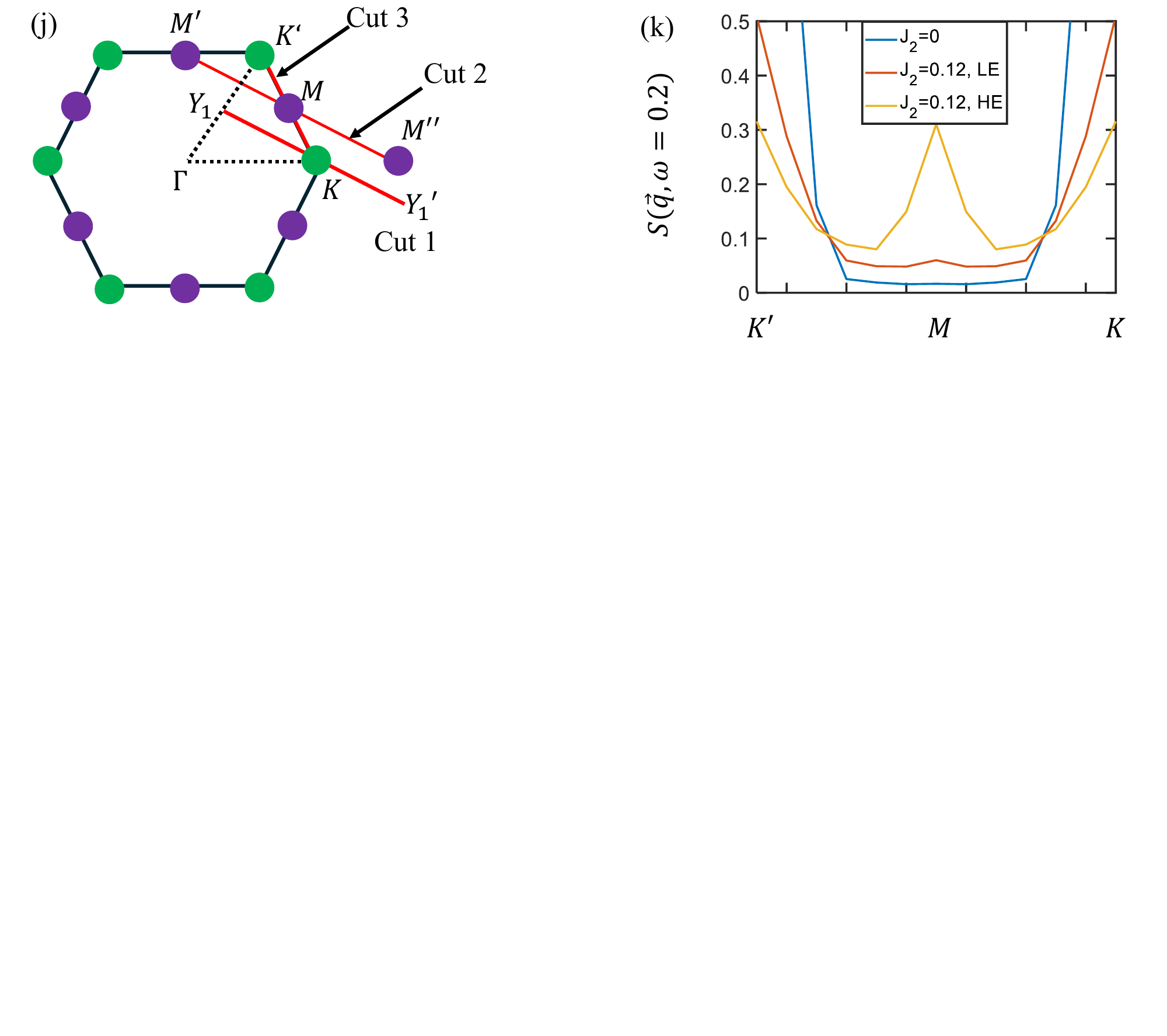}
    \vskip -5.3cm
	\caption{On an XC6-24 cylinder, the DSSF $S(\vec q,\omega)$ along momentum cut 1 (top row, (a)(b)(c)), momentum cut 2 (middle row, (d)(e)(f)), and momentum cut 3 $J_2=0.12$ (bottom row, (g)(h)(i));  for the $J_2=0$ state  (left column, (a)(d)(g)), the LE state at $J_2=0.12$ (middle column, (b)(e)(h)), and the HE state at $J_2=0.12$ (right column, (c)(f)(i)), respectively. (j) Illustrates the three momentum cuts. The color scale has an upper cutoff of 1.3. (k) Along cut 3, the low-energy $S(\vec q,\omega=0.2)$ for the three states.}
	\label{fig:dispersion}
    \vskip -0.5cm
\end{figure}

{\it Dynamical spin structure factor.}--
Using the improved DDMRG algorithm, we computed the DSSF (Eq.~\ref{eq:dssf}) for various states. A broadening factor $\eta=0.1$ was used, and we have checked that the qualitative features of the states are insensitive to the choice of $\eta$~\cite{sm}. Calculations were carried out on XC6-24, XC8-24, and XC9-24 cylinders, keeping a maximum bond dimension of $m$=1800, 3000, and 5000, respectively, using $U(1)$ symmetry. 
To better approximate the 2D system 
with $C_6$ rotational symmetry, we choose data from a central $L_y\times L_y$ region to perform the Fourier transform in Eq.~\ref{eq:dssf}, such that the 
accessible momenta are approximately $C_6$ symmetric. 
Additionally, we also performed a Fourier transform using data from the central half of the system to obtain the dispersive DSSF along several momentum cuts, as shown in Fig.~\ref{fig:dispersion} in the End Matter.

In Fig.~\ref{fig:sqw} we show the low-energy DSSF $S(\vec q,\omega)$ at fixed $\omega$, for the 120$^\circ$ state at $J_2=0$, the LE and HE states at $J_2=0.12$. 
In each panel, the value of $\omega$ was taken to be that which maximizes either $S(\vec q=K,\omega)$ or $S(\vec q=M,\omega)$, whichever lies lower in energy, see Fig.~\ref{fig:sqwline} below. The $J_2=0$ state has clear peaks at the ordering vector $\vec q=K$, corresponding to the 120$^\circ$ magnetic order, as shown in Fig.~\ref{fig:sqw}(a)(b)(c) for XC6-24, XC8-24, and XC9-24 cylinders, respectively.  The LE state at $J_2=0.12$ also has dominant peaks at $\vec q=K$ with a similar amplitude to the $J_2=0$ case, but with a somewhat more diffuse structure around it. The intensity at the $M$ point increases only slightly and does not exhibit a clear peak. 

In contrast, the low-energy DSSF for the HE state at $J_2=0.12$ exhibits conspicuous differences from the previous two states, as shown in Figs.~\ref{fig:sqw}(g)(h)(i). The intensity at the $K$ point is weaker, and there is a clear peak at the $M$ point with comparable intensity, when it is accessible in XC6 and XC8 cylinders. In all cases, the DSSF is much more diffuse along $\vec q=K \rightarrow M$. 
The features in Fig.~\ref{fig:sqw} can be better highlighted after $D_6$-symmetrization and interpolation, provided in the SM~\cite{sm}.

We further compare the DSSF at or near the high symmetry points $\vec q=K$ and $\vec q=M$ in Fig.~\ref{fig:sqwline}. Due to the discrete transverse quantization of momenta, some of the exact high symmetry points are inaccessible on the XC8 or XC9 cylinder~\footnote{On XC8 cylinder, the $\sim K$ point is chosen at a nearby momentum $(1.25\pi,~0.1443\pi)$ where the intensity reaches its maximum. The $\sim M$ point is taken to be $(1.111\pi,~0.385\pi)$ on XC9 cylinder, which is closest to the original $M$ point along the zone boundary. For the HE state on XC9 cylinder, the intensity at $(1.25\pi,~0.257\pi)$ is chosen, since the peak has slightly shifted away from the original $K$ point (Fig.~\ref{fig:sqw}(i)).}.
Another consequence of the cylindrical boundary conditions is that the spectra are not $C_6$ symmetric, especially as is apparent in Fig.~\ref{fig:sqw}, at the $M$ points.
On the other hand, due to translation and inversion symmetries, the intensities at or around the $K$ points are the same. To compare intensities at these two momenta, the $M$ points data in Fig.~\ref{fig:sqwline} have been averaged over the symmetry-related points.

For $J_2=0$, the strongest peak appears at $\vec q =K$. The associated gap, defined by the peak position of $S(K,\omega)$ in Fig.~\ref{fig:sqwline}, decreases quickly as $L_y$ increases, which is consistent with the expected appearance of a Goldstone mode in the 2D limit.
In contrast, the low-energy excitation at $\vec q=M$ is strongly suppressed.
For the LE state at $J_2=0.12$, the dominant peak remains at $\vec q=K$, although with a larger gap than the $J_2=0$ case. The spectral weight at $\vec q=M$ is somewhat larger than for $J_2=0$ and moves to lower frequency with increasing $L_y$, but remains much smaller at low energy than the $K$ point and never exhibits a well-defined peak frequency.
For the HE state, the low-energy excitations at the $K$ and $M$ points are comparable in strength, 
and the gaps at both momenta appear to decrease as $L_y$ increases. We have also verified that these characteristics of the HE and LE states behave similarly on YC6-24 cylinders~\cite{sm}.

\begin{figure}[t]
	\includegraphics[width=1.0\linewidth]{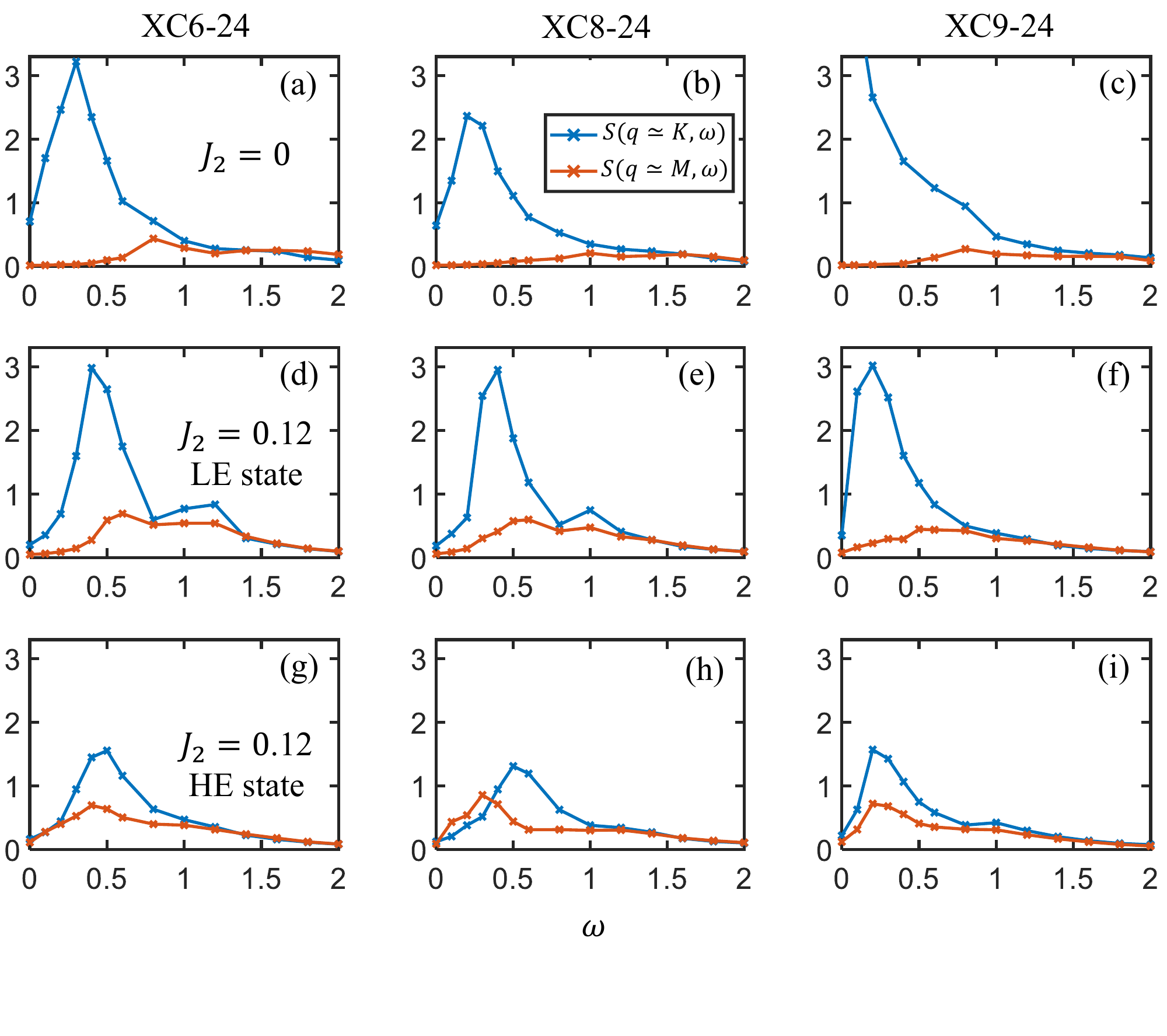}
    \vskip -0.9cm
	\caption{The DSSF $S(\vec q,\omega)$ at or around high symmetry points $\vec q=K=(4\pi/3,0)$ and $\vec q=M=(\pi,\pi/\sqrt3)$ after symmetrization (see text), for the 120$^\circ$ state at $J_2=0$ (top row, (a)(b)(c)), the LE state at $J_2=0.12$ (middle row, (d)(e)(f)), and the HE state at $J_2=0.12$ (bottom row, (g)(h)(i)). Simulations 
    were carried out on XC6-24 (left column, (a)(d)(g)), XC8-24 (middle column, (b)(e)(h)), and XC9-24 (right column, (c)(f)(i)), respectively.}
	\label{fig:sqwline}
\end{figure}

Having established the main characteristics of the DSSF in different cases, we now compare them with the theoretical expectations for various candidate phases. According to theoretical predictions for the gapped $\mathbb{Z}_2$ QSL~\cite{z2-theory-ashvin,z2-theory-sachdev}, the spinon dispersion should have a minimum at the $K$ point, which leads to a magnon minimum at $2K$ (equivalent to $K$). 
Within this framework, the system can undergo a continuous phase transition from the 120$^\circ$ state to the gapped $\mathbb{Z}_2$ QSL accompanied by a gap opening at $K$.
Although the LE state identified on finite cylinders is compatible with this scenario, the interpretation is not unique.
For finite cylinder circumference, the 120$^\circ$ state at $J_2=0$ also exhibits a finite gap, as shown in Fig.~\ref{fig:sqwline}(a-c) (the gap in XC9-24 is around 0.08).
As $J_2$ increases and magnetic order is weakened, the gap on a finite cylinder correspondingly grows; this behavior does not preclude the presence of magnetic order in the 2D limit. Indeed, finite-size scaling of the magnetic order extrapolates suggestively to a tiny but non-zero value in the 2D limit~\cite{sm}. This is accompanied by a decrease in the gap as $L_y$ widens, although it is unclear if it will close in 2D. Based on our results on cylinders up to width XC9, we cannot decisively determine whether the LE state evolves into a gapped $\mathbb{Z}_2$ QSL or instead reflects a weakly magnetically ordered state in the 2D limit. 

On the other hand, the low-energy effective field theory of a $U(1)$-Dirac QSL on the triangular lattice is quantum electrodynamics in 2+1 dimensions (QED3)~\cite{dirac-theory,dirac-theory2,qed3-lauchli}, which predicts gapless spin-triplet excitations at both the $M$ and $K$ points of the Brillouin zone. The HE state observed in our calculations appears to be consistent with this expectation.
It displays low-energy spectral weight at both $M$ and $K$ with comparable intensities, and its excitation gap decreases as $L_y$ increases.
Similar signatures of a Dirac QSL have also been reported in previous numerical studies~\cite{j1j2-becca-prx,j1j2-moore,j1j2-imada,j1j2-pollmann,j1j2-pollmann-2025,qed3-lauchli}. 
In contrast, the LE state deviates from the Dirac QSL predictions, in that the low-energy triplet spectral weight is strongly suppressed at the $M$ point compared to the $K$ point. This appears to be inconsistent with its being a $U(1)$ Dirac QSL in the 2D limit.

{\it Discussion.}--
While the HE state is most probably a gapless Dirac QSL in 2D,  we are currently unable to convincingly determine the precise nature of the LE state, nor 
are we able to conclude which state will evolve into the equilibrium ground state in the 2D limit. Addressing these questions will require even wider systems, particularly those designed so that the allowed crystal momenta include both the $K$ and $M$ points.
It will also require clever finite-size scaling analysis of the magnetic order parameter of the 120$^\circ$ state and the spin-triplet excitation gap, 
to distinguish between a weakly magnetically ordered state and a gapped $\mathbb{Z}_2$ QSL.
Note that the existence of two distinct, nearly degenerate QSL states has also been inferred in studies of the kagome-lattice AFM~\cite{kagome-eduardo,z2-theory-sachdev}.

\textit{Data availability}: 
The data used to generate the figures are deposited in Zenodo~\footnote{\href{https://doi.org/10.5281/zenodo.18535976}{https://doi.org/10.5281/zenodo.18535976}}.

\begin{acknowledgments}
\emph{Acknowledgments.}---
We acknowledge Leon Balents, Cenke Xu, and Stephen Shenker for helpful discussions.
S.J., S.A.K., and H.C.J. are supported by the U.S. Department of Energy (DOE), Office of Science, Basic Energy Sciences, Materials Sciences and Engineering Division, under contract DE-AC02-76SF00515. S.J. and the DDMRG calculations were supported by the Department of Energy, Laboratory Directed Research and Development program at SLAC National Accelerator Laboratory, under contract DE-AC02-76SF00515, with partial support for H.C.J..
SRW is supported by the NSF under DMR-2412638. 
Calculations were performed on resources of the National Energy Research Scientific Computing Center, supported by the U.S. Department of Energy under contract DEAC02-05CH11231.
\end{acknowledgments}

\clearpage
\onecolumngrid
\begin{center}
\vskip 0.15cm
{\large\bf End Matter}
\end{center}
\vskip 0.15cm
\twocolumngrid

\begin{figure}[htb]
  \includegraphics[width= 1.0\linewidth]{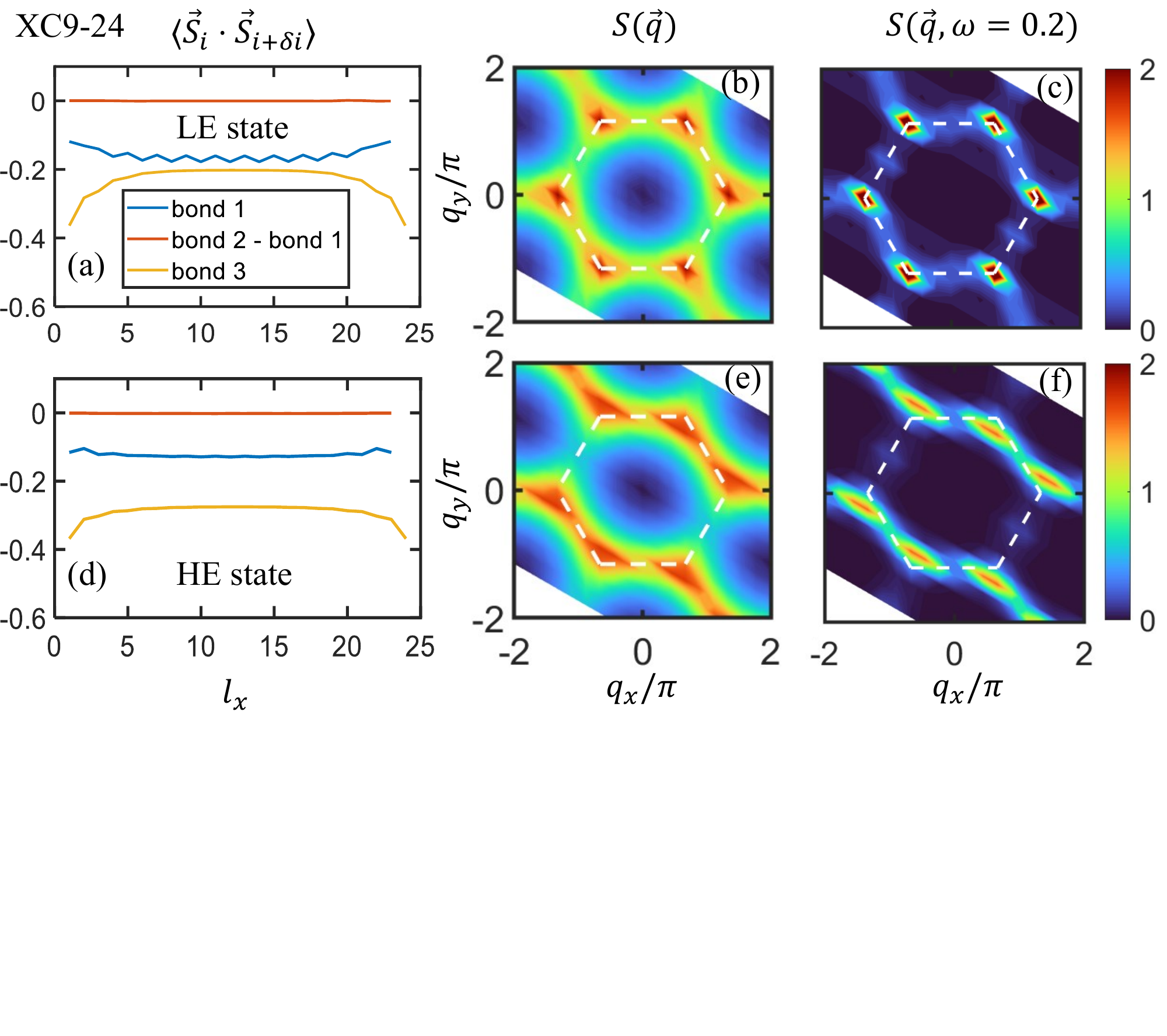}
  \vskip -2.6cm
  \caption{Correlation functions in the LE (top panels) and HE state (lower panels) in a XC9-24 cylinder, with $J_2=0.12$.
  Panels (a) and (d) plot the strength of the spin-spin correlation $\langle \vec S_i\cdot \vec S_{i+\delta i}\rangle$ on three types of NN bonds (defined in Fig.~\ref{fig:phd}(a) in the main text), averaged over the circumferential direction and plotted versus the position along the cylinder $l_x$. Panels (b) and (e) show the equal-time spin structural factor $S(\vec q)$. (c) and (f) show the  
  the low-energy DSSF $S(\vec q,\omega=0.2)$.}
  \label{fig:xc9}
\end{figure}

{\it Differences between the LE and HE state on the XC9 cylinder.}--
Here we summarize some further  properties of the LE state and the HE state on XC9-24 cylinders, with a focus on the evidence against assigning a topological interpretation to the existence of a metastable HE state.
On the basis of an analysis~\cite{yao,j1j2-zhu} of the quantum dimer model of a $\mathbb{Z}_2$ QSL, one expects that on a cylinder of finite width with $L_y$ odd, one should find a state with broken translational symmetry (i.e. columnar dimer order with strength) that decreases exponentially with $L_y$. 
In fact, as shown in Fig.~\ref{fig:xc9}(a), 
we do see a columnar order, i.e. pronounced period 2 spatial oscillations in the strength of the spin correlations on bond-types 1 and 2 in the LE state, consistent with the above $\mathbb{Z}_2$ QSL description. However, 
to the extent that there are any such oscillations in the HE state (Fig.~\ref{fig:xc9}(b)), they are substantially weaker - again inconsistent with the two states being only topologically distinct.

For both the equal-time structural factor $S(\vec q)$ and the low-energy dynamical structural factor $S(\vec q,\omega=0.2)$, the LE state is qualitatively similar to the ordered state at $J_2=0$ with a dominate peak only at $\vec q=K$. The HE state has a more diffuse structure factor from $K$ to $M$, similar to that on even-width cylinders and more consistent with a gapless Dirac QSL.
We note that that there is an interesting difference with the situation in the even leg cylinders we have studied in that the HE state appears to be less robustly metastable: it remains metastable up to $m=2400$, but evolves towards the LE state at 
larger $m$. The DMRG results shown in the figure for the LE states and the DDMRG results for both states are at $m=4000$.

\begin{figure}[hbt]
	\includegraphics[width= 1.0\linewidth]{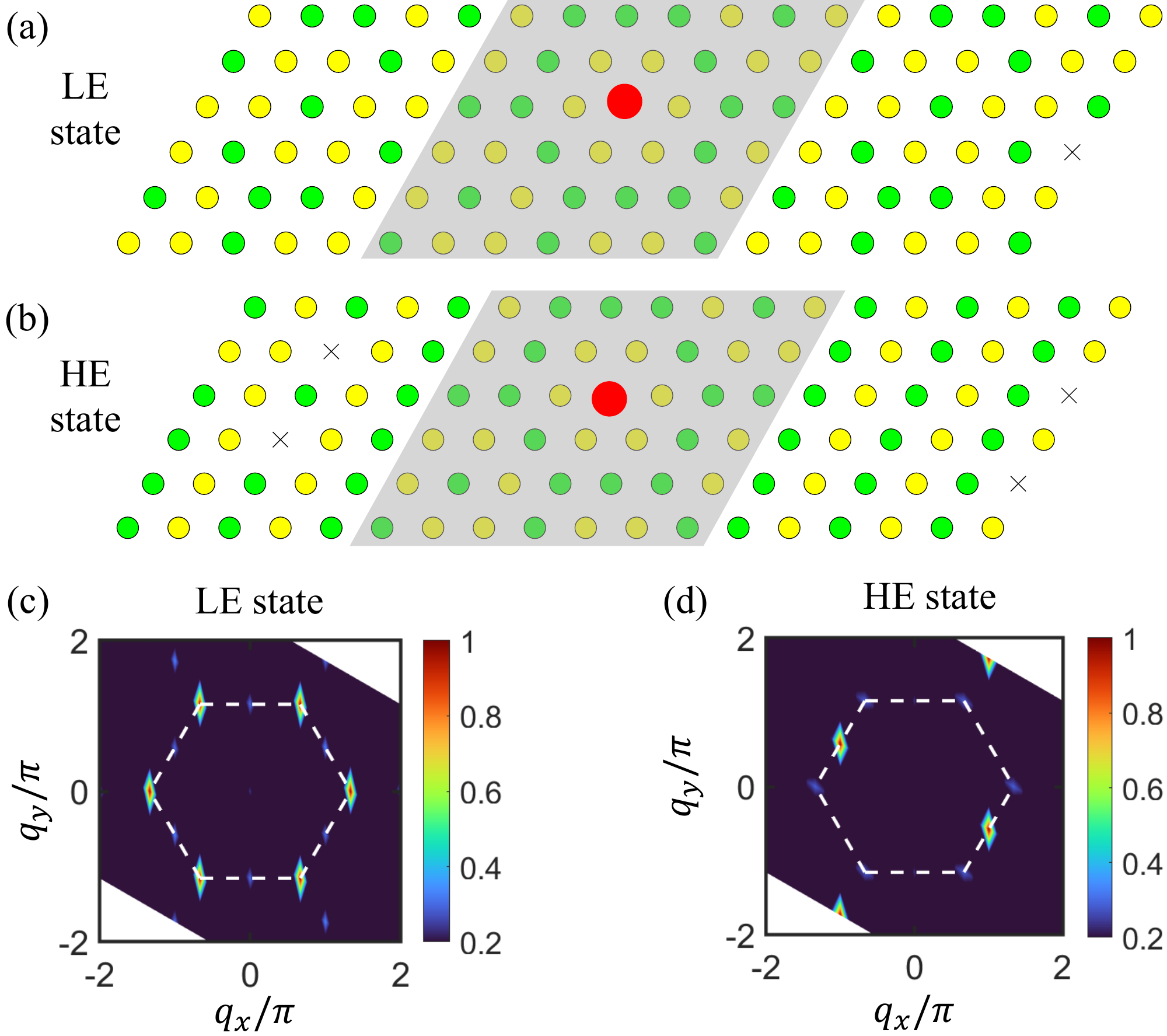}
	\caption{Sign of the spin-spin correlation function $\langle \vec{S}_{\vec{r}_0} \cdot \vec{S}_{\vec{r}}\rangle/ |\langle \vec{S}_{\vec{r}_0} \cdot \vec{S}_{\vec{r}}\rangle|$ for (a) the LE state and (b) the HE state at $J_2=0.12$ on a XC6-24 cylinder at $m=2400$, with three columns subtracted on each edge. The reference site $\vec r_0$ is marked by the red circle. Green dots and yellow dots denote plus and minus signs at site $\vec r$, respectively. Those sites with a correlation amplitude smaller than $5\times 10^{-5}$ are excluded and marked by a cross, since their signs are susceptible to errors and can not be definitely determined.
    Outside the shaded vicinity of $\vec r_0$, the patterns of the correlation are different for the two states.
    The Fourier transform of the sign of the spin-spin correlation functions: $\tilde{S}(\vec{q})=\sum_{\vec{r}} e^{-i\vec{q} \cdot \vec{r}}  \langle \vec{S}_{\vec{r}_0} \cdot \vec{S}_{\vec{r}}\rangle/ |\langle \vec{S}_{\vec{r}_0} \cdot \vec{S}_{\vec{r}}\rangle|$, for (c) the LE state and (d) the HE state. The color scale is renormalized and has a lower cutoff of 0.2 to filter out the fuzzy background due to contributions from short-ranged correlations.}
	\label{fig:ftsign}
\end{figure}

{\it Structure of the sign of the equal-time spin-spin correlation function}---
Here we further analyze the equal-time spin-spin correlation function, focusing on its sign pattern and ignoring its amplitude, plotted in Fig.~\ref{fig:ftsign}(a) and (b) for the LE and HE states, respectively. While the short-ranged correlations marked by the shaded region are identical for the two states, the longer-ranged correlations are significantly different. The LE state mostly follows a three-sublattice pattern, similar to the 120$^\circ$ state, while the HE state exhibits a striped pattern.  

We also look at the structural factor of the sign of the correlation functions, as shown in Fig.~\ref{fig:ftsign}(c) and (d). They also exhibit a significant distinction: the LE state resembles the 120$^\circ$ state with peaks at $\vec q=K$, while the HE state resembles the striped state with peaks at $\vec q=M$. This contrasts with the ordinary $S(\vec q)$ in Fig.~\ref{fig:phd}(f) and (g), where the difference is less significant, since it is dominated by short-range correlations.
Similar distinction in the sign pattern between the two states has also been observed in XC8 and XC9 cylinders, suggesting their different natures.

{\it Approximation in DDMRG and related issues.}---
While ground-state DMRG is variational, DDMRG is usually implemented in a way that is not. 
When solving the correction-vector equation (Eq.~\ref{eq:corr_vec}), one uses real arithmetic to solve the imaginary part $|Y\rangle$ first via $[(E_0+\omega-H)^2+\eta^2]|Y\rangle=-\eta S|0\rangle$, which can then be used to generate the real part $|X\rangle=(H-E_0-\omega)|Y\rangle/\eta$~\cite{ddmrg-jeckelmann}. 
Since the matrix product operator of $H^2$ has a large bond dimension, the standard approach is to apply $H$ twice in the same basis~\cite{ddmrg-jeckelmann}, i.e., approximating $H^2$ by $(H^{\rm proj})^2$ with $H^{\rm proj}$ being the reduced superblock Hamiltonian after projecting on the two-site basis. 
The accuracy of this approximation relies on the basis representing the intermediate state $H^{\rm proj}|Y\rangle$ accurately, as well as the initial $|Y\rangle$ and final states $S|0\rangle$~\cite{ddmrg-jeckelmann}. 
In practice, a truncation on the intermediate state always occurs during DDMRG, which makes the calculation non-variational. 
This means that the iterative linear solver can have slow convergence or ``fake" convergence for the \emph{projected} problem when the true (full-space) residual remains large, leading to inaccurate spectra at limited bond dimension. This issue is exacerbated by the fact that the standard state-averaged implementation in DDMRG must accommodate four distinct targets---$|0\rangle$, $S|0\rangle$, $|X\rangle$, and $|Y\rangle$---within a single truncated MPS basis; with limited $m$, the resulting truncation can be substantial and significantly distort the reduced linear problem.

{\it Dynamical spin structure factor along momentum cuts.}---
To supplement the DSSF at specific low-energy (Fig.~\ref{fig:sqw}) and at specific momenta (Fig.~\ref{fig:sqwline}), we plot the DSSF in XC6-24 cylinders along several momentum cuts.  
To better see the dispersive features, we calculate the DSSF in Eq.~\ref{eq:dssf} with $\vec r_j$ being the reference site in the center of the cylinder and $\vec r_i$ running over the central half of the system. This allows us to access more momentum points along the length of the cylinder (compared with using data from the central $L_y\times L_y$ region), such as cut 1 and cut 2 in Fig.~\ref{fig:dispersion}(j).
After that, we perform a $D_6$ symmetrization of the DSSF and interpolate the data between the accessible momenta. Note that the DSSF obtained this way will have a minor quantitative difference from running both $\vec r_j$ and $\vec r_i$ over the $L_y\times L_y$ region as in the previous figures, but the qualitative features remain the same.

For both the $J_2=0$ and the LE state at $J_2=0.12$, one can see a clear suppression of low-energy excitations at the $M$ point. The LE state has a somewhat more diffuse structure at higher energies, but overall it appears qualitatively similar to the $J_2=0$ state. On the other hand, for the HE state at $J_2=0.12$, low-energy modes emerge at $M$ with intensities comparable to those of $K$. In Fig.~\ref{fig:dispersion}(k), we plot the intensity of DSSF along the momentum cut 3 at an energy cut $\omega=0.2$, where one can see a clear peak at $M$ that is absent in the other two states, signifying their different natures. 

\bibliography{ref}
\end{document}


\title{Supplemental materials: Competing states in the $S=1/2$ $J_1$-$J_2$ Heisenberg model on the triangular lattice: a dynamical density-matrix renormalization group study}

\begin{CJK*}{UTF8}{}
\author{Shengtao Jiang (\CJKfamily{gbsn}蒋晟韬)}
\affiliation{Stanford Institute for Materials and Energy Sciences, SLAC National Accelerator Laboratory and Stanford University, Menlo Park, CA 94025, USA}

\author{Steven R. White}
\affiliation{Department of Physics and Astronomy, University of California, Irvine, California 92697, USA}

\author{Steven A. Kivelson}
\affiliation{Department of Physics, Stanford University, Stanford, CA 94305, USA}

\author{Hong-Chen Jiang}
\affiliation{Stanford Institute for Materials and Energy Sciences, SLAC National Accelerator Laboratory and Stanford University, Menlo Park, CA 94025, USA}

\date{\today}
\maketitle
\end{CJK*}

\section{Review of the original Dynamical density-matrix renormalization group algorithm}
In this section, we briefly review the original dynamical density-matrix renormalization group (DDMRG) method~\cite{ddmrg-jeckelmann}, highlighting the approximations and limitations that motivate the improvements introduced in the next section.
While DDMRG can be used to calculate general dynamical correlation functions, the goal in this paper is to compute the dynamical spin structure factor (DSSF), defined as:
\begin{equation}
\begin{split}
S(\vec q,\omega)=&\frac{-1}{\pi N}\sum_{\substack{a,b=1\\\alpha=x,y,z}}^{N}e^{i\vec{q}\cdot(\vec{r}_a-\vec{r}_b)}\times\\
&\text{Im}\langle 0|S^\alpha_{\vec{r}_a}\frac{1}{E_0+\omega-H+i\eta}S^\alpha_{\vec{r}_b}|0\rangle,
\label{eq:dssf}
\end{split}
\end{equation}
where $|0\rangle$ is the ground state obtained by DMRG with energy $E_0$, and $\eta$ is the chosen broadening factor. 
We define the correction vector $|c^\alpha_{\vec{r}_b}(\omega)\rangle$ that satisfies:
\begin{equation}
(E_0+\omega-H+i\eta)|c^\alpha_{\vec{r}_b}(\omega)\rangle=S^\alpha_{\vec{r}_b}|0\rangle, 
\end{equation}
which can be used to compute $S(\vec q,\omega)$:
\begin{equation}
S(\vec q,\omega)=\frac{-1}{\pi N}\sum_{\substack{a,b=1\\\alpha=x,y,z}}^{N}e^{i\vec{q}\cdot(\vec{r}_a-\vec{r}_b)}\times\text{Im}\langle 0|S^\alpha_{\vec{r}_a}|c^\alpha_{\vec{r}_b}(\omega)\rangle.
\end{equation}

The complex correction vector is separated into real and imaginary parts: $|c^\alpha_{\vec{r}_b}(\omega)\rangle=|X^\alpha_{\vec{r}_b}(\omega)\rangle+i|Y^\alpha_{\vec{r}_b}(\omega)\rangle$, so that it can be solved with real arithmetic. Its imaginary part satisfies:
\begin{equation}
    [(E_0+\omega-H)^2+\eta^2]|Y^\alpha_{\vec{r}_b}(\omega)\rangle=-\eta S^\alpha_{\vec{r}_b}|0\rangle, 
    \label{eq:ddmrg}
\end{equation}
based on which the real part can also be calculated:
\begin{equation}
    |X^\alpha_{\vec{r}_b}(\omega)\rangle=\frac{H-E_0-\omega}{\eta}|Y^\alpha_{\vec{r}_b}(\omega)\rangle.
\label{eq:X_from_Y}
\end{equation}
Generally, multiple calculations are needed for each reference site $\vec r_j$ and spin component $\alpha$. If the ground state and Hamiltonian possess translational and spin rotational symmetry, then this can be simplified, and only one calculation is needed for each $\omega$. From now on, we will drop the subscript $\vec r_j$ and superscript $\alpha$ to simplify notation.

The dynamical-DMRG algorithm~\cite{ddmrg-jeckelmann} solves Eq.~\ref{eq:ddmrg} using the same sweep scheme as the DMRG algorithm~\cite{White1992,White1993}, replacing the original eigenvalue problem with the linear problem in Eq.~\ref{eq:ddmrg}.
At each step of the sweep, one solves a reduced linear problem by projecting the original problem (Eq.~\ref{eq:ddmrg}) on a reduced two-site basis. This is similar to DMRG where one diagonalizes the effective superblock Hamiltonian $H^{\rm eff}$ formed by $H^{\rm eff}=PHP$, where $P$ is the operator projecting on the two-site basis. However, since only $H$ is accessible as an MPO but not $H^2$, one has to introduce another approximation in DDMRG:
\begin{equation}
    P(H-E_0-\omega)^2P\approx(H^{\rm eff}-E_0-\omega)^2=(PHP-E_0-\omega)^2,
    \label{eq:approxddmrg}
\end{equation}
and this approximated reduced linear equation is solved using the conjugate gradient (CG) method. 
To minimize the effect of this approximation, $|X(\omega)\rangle$ (Eq.~\ref{eq:X_from_Y}) needs to be state-averaged with $|Y(\omega)\rangle$, $|0\rangle$ and $S|0\rangle$ in DDMRG. In this way, all these states share a common basis based on which the reduced linear problem is constructed. 
If there is no truncation on $|X(\omega)\rangle$, then the approximation error in Eq.~\ref{eq:approxddmrg} vanishes.
However, since truncation generally occurs, DDMRG algorithm is in-principle non-variational and can have slow or incorrect convergence. 
This is because the approximation (Eq.~\ref{eq:approxddmrg}) can distort the original linear problem (Eq.~\ref{eq:ddmrg}), such that the residual measured within the reduced basis may appear small and converged, while the true residual in the full Hilbert space remains large.
This issue can be exacerbated when the bond dimension is insufficient and a heavy truncation happens. 
This is the primary source of instability that will be addressed in the next section.

The advantage of DDMRG over time-evolution-based methods is in obtaining low-frequency (small $\omega$) dynamical correlation functions~\cite{ddmrg++}. In time evolution, the entanglement entropy grows with time, and one has to apply heavy truncation or stop after a certain time. As a result, the precision of the low-frequency correlations suffers more significantly from truncation, since they involve the long time signal.
In contrast, DDMRG calculates the dynamical correlations directly in frequency space.
Low-energy excited states typically exhibit lower entanglement entropy, making them more accessible in DDMRG.
On the other hand, the higher-energy excited states have larger entanglement entropy and are more difficult to obtain. However, because the excitation is usually made of a local distortion, such as a spin flip or an electron removal, there is an upper bound on the excitation energy. 
As a result, although the relative error can be bigger for the high-energy excited states, their overall weight is also smaller and are less important in the spectral function.

\section{Improved Dynamical density-matrix renormalization group algorithm}
In this section, we introduce several modifications to the original DDMRG algorithm, such that the correct qualitative features can be captured at a modest bond dimension with stable convergence.
The modifications are:  (1) use a separate MPS to store the ground state $|0\rangle$ and $S|0\rangle$ (following Ref.~\cite{ddmrg++}), mix the latter with a weight $\sqrt \eta$ in the state-averaged MPS, to mitigate the projection error when constructing the reduced linear problem (Eq.~\ref{eq:ddmrg}); (2) impose a weight $\eta$ on $|X(\omega)\rangle$ to de-emphasize it during state-averaging, which can preserve the accuracy of $|Y(\omega)\rangle$ that can help stabilize the algorithm when the bond dimension is relatively small. In this way, three states ($\sqrt \eta S^\alpha_{\vec{r}_b}|0\rangle$, $\eta|X(\omega)\rangle$, and $|Y(\omega)\rangle$) are state-averaged in one MPS, with the latter two being optimized during DDMRG.

In conventional DDMRG, one needs to state-averaging $|Y(\omega)\rangle$, $|X(\omega)\rangle$, $|0\rangle$, and $S|0\rangle$ in one MPS.  
When the bond dimension is small, it can be hard to represent all four states accurately simultaneously, resulting in two problems. 
The first one is observed in Ref.~\cite{ddmrg++} that the compressed $|0\rangle$ and $S|0\rangle$ in the state-averaged MPS can deviate significantly from the uncompressed ones, and require further optimization during DDMRG. To fix this, they propose using two separate MPSs, one representing $|0\rangle$ and $S|0\rangle$ that are held fixed during DDMRG, while the other represents $|X(\omega)\rangle$ and $|Y(\omega)\rangle$. The former is projected on the basis of the latter to form the reduced linear problem in Eq.~\ref{eq:ddmrg}. 

We further observe that the basis of $|X(\omega)\rangle$ and $|Y(\omega)\rangle$ can be quite different from that of $S|0\rangle$, especially when the excitation energy $\omega$ is high. In such a case, the error can be substantial when projecting $S|0\rangle$ on the basis of $|X(\omega)\rangle$ and $|Y(\omega)\rangle$. 
Therefore, it is essential to also mix $S|0\rangle$ with $|X(\omega)\rangle$ and $|Y(\omega)\rangle$ in the state-averaged MPS during the DDMRG sweep to reduce the projection error.

As described in more detail in the previous section, another potential issue concerns the approximation in Eq.~\ref{eq:approxddmrg}. When the bond dimension is limited, one needs to heavily truncate $|X(\omega)\rangle$ and $|Y(\omega)\rangle$, which can result in difficulty converging or converging to an incorrect solution. The key question is how to allocate the bond dimension budget between  $|X(\omega)\rangle$, $|Y(\omega)\rangle$, and $S|0\rangle$ such that the correct dynamical features can be captured with stable convergence at a modest bond dimension. 

Empirically, we have found that enhancing the weight of $|Y(\omega)\rangle$ at the sacrifice of the other two can help achieve this goal. Specifically, we impose a weight of $\eta$ and $\sqrt\eta$ on $|X(\omega)\rangle$ and $S|0\rangle$, respectively, during state-averaging.
Although this may appear counterintuitive, as the approximation error arises because of truncation on $|X(\omega)\rangle$, one would expect that enhancing its weight can help reduce the error.
The reason is that because $|X(\omega)\rangle$ is generated from $|Y(\omega)\rangle$ (Eq.~\ref{eq:X_from_Y}), it is more important to represent $|Y(\omega)\rangle$ accurately first, without which the correct $|X(\omega)\rangle$ cannot be generated. 
From another perspective, it is also important to avoid a heavy truncation to preserve the optimized $|Y(\omega)\rangle$ after CG iterations. 
The weights $\eta$ and $\sqrt \eta$ on them are empirical parameters that appear to set the appropriate balance between the states. Because of these reduced weights, the entanglement entropy from the reduced density matrix at state-averaging reduces, resulting in a smaller truncation error. 

\begin{figure}[ht]
  \includegraphics[width= 0.65\linewidth]{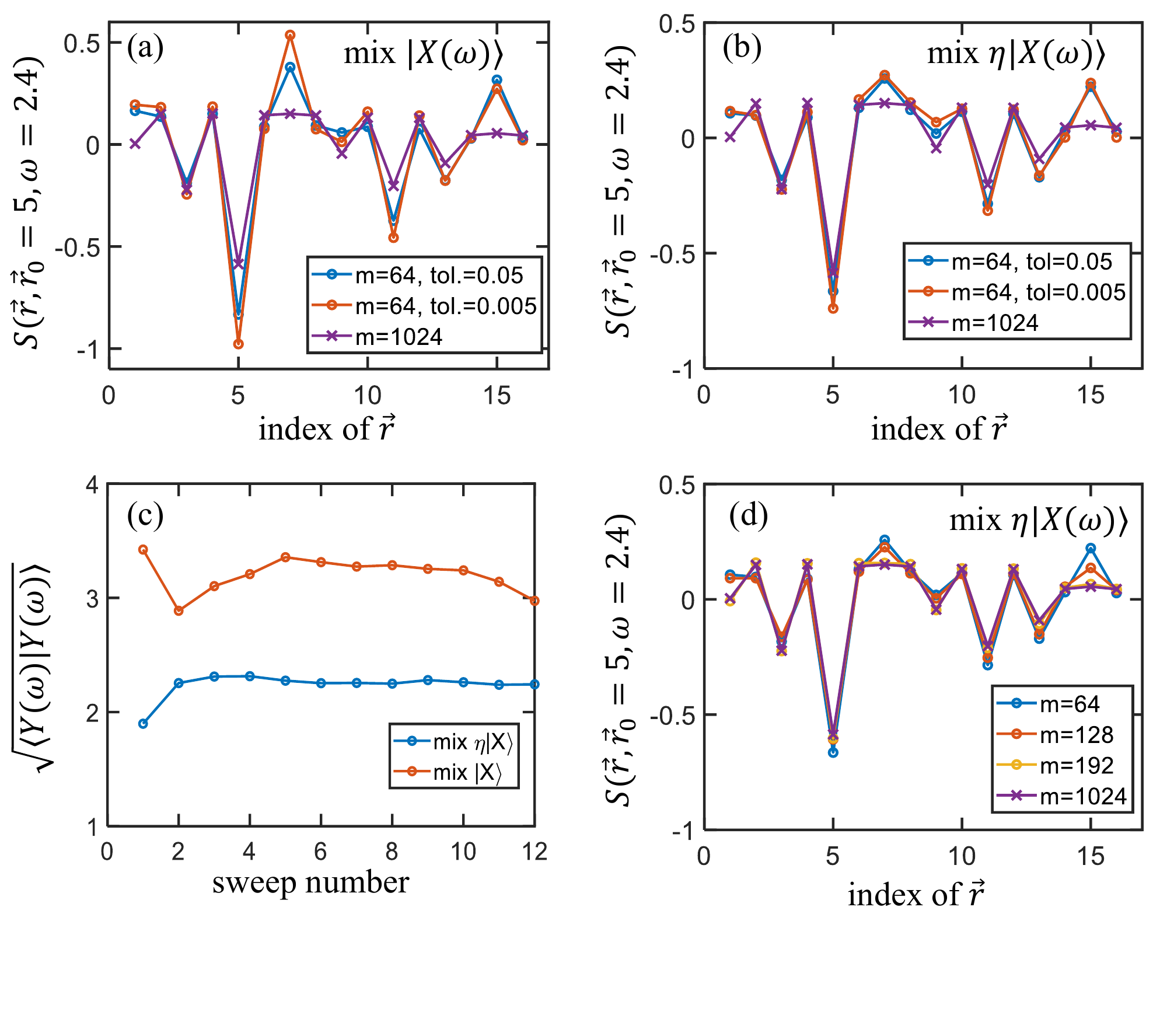}
  \vskip -1.5cm
  \caption{(a) The Dynamical spin correlation function $S(\vec r,\vec r_0,\omega)$ (Eq.~\ref{eq:sromega}) of the Heisenberg model on a $4\times 8$ square cylinder at $\omega=2.4$ using $\eta=0.05$, at a bond dimension $m=64$. $\vec r$ iterates over the central 4x4 region and its index equals $r_y+4(r_x-3)$. The $m=1024$ data can be considered as exact. Here we mix $|X(\omega)\rangle$ in the state-averaged MPS, and the results are unstable with respect to the number of conjugate gradient (CG) iterations controlled by the tolerance. (b) Same as (a) but mixing $\eta|X(\omega)\rangle$ instead of $|X(\omega)\rangle$ when state-averaging. The algorithm is more stable and the results improve at $m=64$. (c) The norm of $|Y(\omega)\rangle$ versus sweep number, which has a better convergence behavior when mixing  $\eta|X(\omega)\rangle$ (both using a CG tolerance of 0.05, the norm of the exact $|Y(\omega)\rangle$ is around 1.9.) (d) $S(\vec r,\vec r_0,\omega)$ at different $m$, obtained with mixing $\eta|X(\omega)\rangle$ in the algorithm.}
  \label{fig:convg}
\end{figure}

The improved DDMRG method is benchmarked using the Heisenberg model on a $4\times 8$ square cylinder, with a ground state bond dimension $m=64$ (truncation error $\sim O(10^{-6})$). We directly compared the dynamical spin-spin correlation function $S(\vec r,\omega)$:
\begin{equation}
S(\vec r,\vec r_0,\omega)=\text{Im}\langle 0|S^z_{\vec{r}}\frac{1}{E_0+\omega-H+i\eta}S^z_{\vec{r}_0}|0\rangle,
\label{eq:sromega}
\end{equation}
with $\vec{r}_0$ being the reference site at $(r_x=4,r_y=1)$.
We choose a relatively challenging case with $\omega=2.4J, \eta=0.05J$ (DDMRG calculation generally becomes more difficult at a higher excitation energy $\omega$ and a finer resolution $\eta$).
The results using the original DDMRG method that mixes $|X(\omega)\rangle$ is shown in Fig.~\ref{fig:alg}(a), while those obtained with the improved DDMRG method that mixes $\eta|X(\omega)\rangle$ is shown in Fig.~\ref{fig:alg}(b), which are closer to the exact one at this small bond dimension $m=64$.
The improved DDMRG is also more stable with respect to the number of CG iterations, controlled by the chosen tolerance (tol.), with a maximum of 30 iterations per step.
It also has a more stable convergence behavior, manifested in the evolution of  $\langle Y(\omega)|Y(\omega)\rangle$ with sweeps (Fig.~\ref{fig:alg}(c). 
For the improved DDMRG, we also compare the $S(\vec r,\vec r_0,\omega)$ at different bond dimension $m$, shown in Fig.~\ref{fig:alg}(d). 
For this challenging case, the results at the smallest bond dimension $m=64$ (truncation error $\sim O(10^{-4})$) already captures the correct qualitative features, and are close to the exact ones at a quantitative level, demonstrating the effectiveness of the improved DDMRG method.

\section{Benchmark the improved DDMRG method on square-lattice Heisenberg AFM}

\begin{figure}[ht]
  \includegraphics[width= 0.7\linewidth]{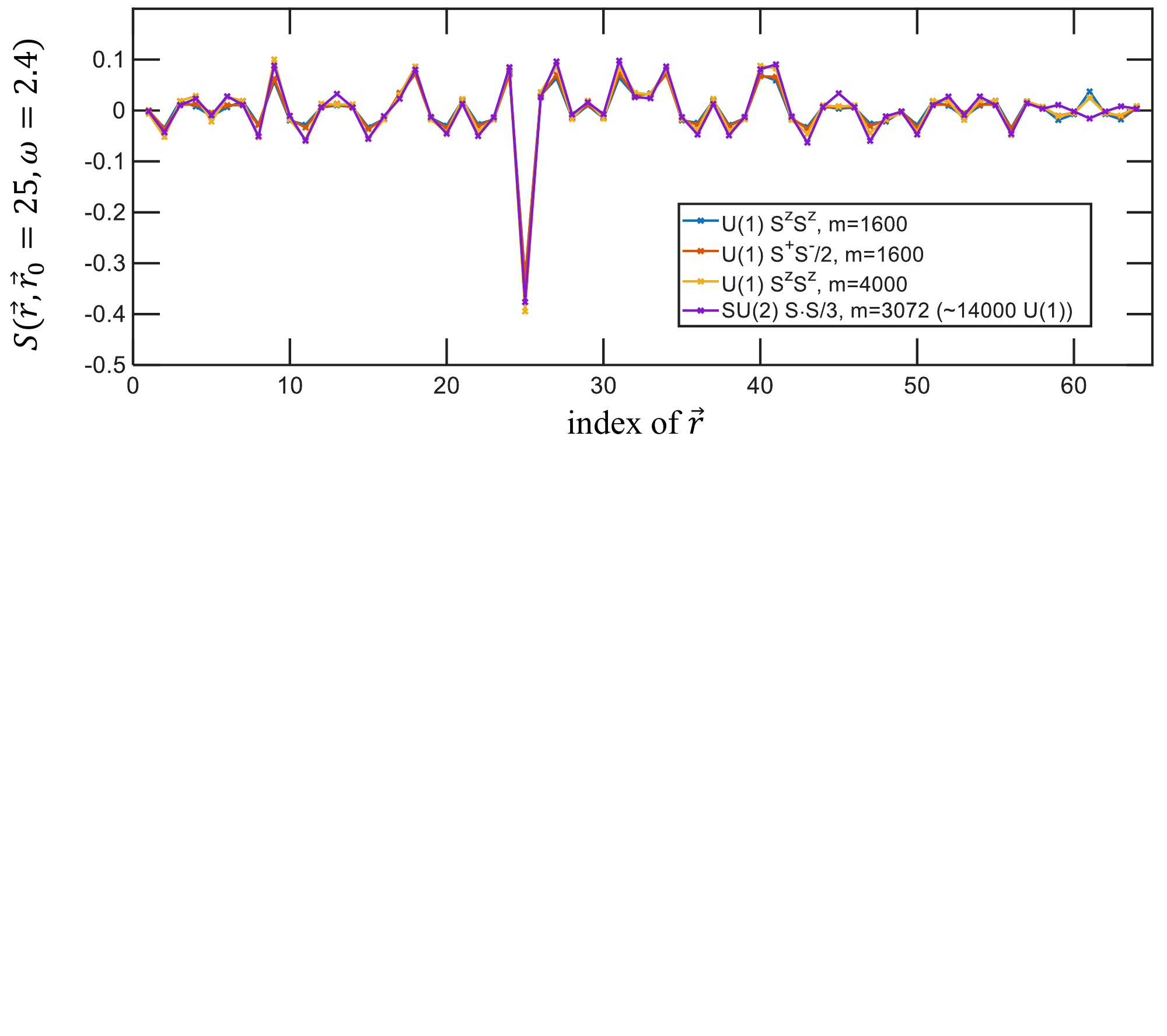}
  \vskip -6.7cm
  \caption{The dynamical spin correlation function $S(\vec r,\vec r_0=25,\omega=2.4)$ (Eq.~\ref{eq:sromega}) of the Heisenberg model on a $8\times16$ square cylinder using $\eta=0.1$, at different bond dimension $m$. $\vec r$ iterates over the central $8\times 8$ region and its index equals $r_y+8(r_x-5)$.}
  \label{fig:srw_square}
\end{figure}

To further benchmark our improved DDMRG method in larger 2D systems, we simulate the $S(\vec q,\omega)$ for the square-lattice Heisenberg antiferromagnet (AFM) on an $8\times16$ cylinder. We use a ground state with $m=800$ that has a truncation error $\sim O(10^{-5})$. We first check the convergence of our DDMRG calculation with respect to bond dimension $m$, at a relatively challenging case with $\omega=2.4$ and $\eta=0.1$. As shown in Fig.~\ref{fig:srw_square}, the dynamical spin-spin correlation function is already well captured at $m=1600$, with little notable differences compared with $m=4000$. It also possesses $SU(2$) symmetry as the $S^zS^z$ correlator is identical to the $0.5S^+S^-$ correlator. The $U(1)$ DDMRG also agrees well with a $SU(2)$ DDMRG at an even larger bond dimension.

\begin{figure}[ht]
  \includegraphics[width= 0.70\linewidth]{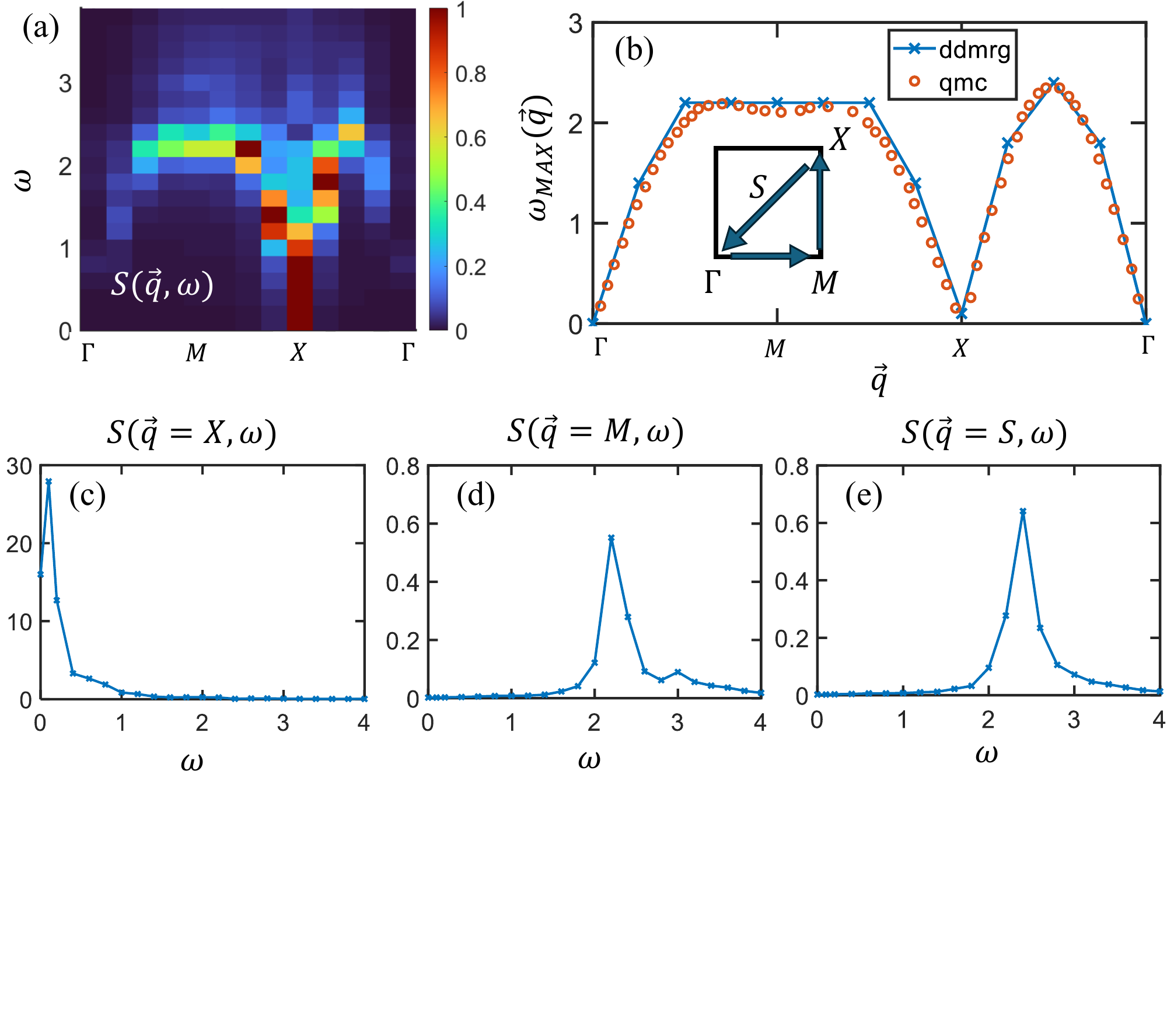}
  \vskip -3.2cm
  \caption{DSSF $S(\vec q,\omega)$ of the square-lattice Heisenberg AFM on $8\times16$ cylinder using a broadening $\eta=0.1$, at bond dimension $m=1600$. Along the high-symmetry path indicated in the inset of (b): (a) the DSSF $S(\vec q, \omega)$ with an upper cutoff of 1 in the color scale, and (b) the peak energy with the biggest intensity $\omega_{\text{MAX}}(\vec q)$. The red dots are extracted from a quantum Monte Carlo simulation in Ref.~\cite{square-qmc}. The DSSF at high symmetry points (a) $X=(\pi,\pi)$, (b) $M=(\pi,0)$ and (c) $S=(\pi/2,\pi/2)$.}
  \label{fig:sqw_square}
\end{figure}

The $S(\vec q,\omega)$ obtained by our improved DDMRG method is shown in Fig.~\ref{fig:sqw_square}. The results are in close agreement with sign-problem-free quantum Monte Carlo (QMC) simulations~\cite{square-qmc}, carried out in a $48\times48$ cluster. In particular, as shown in Fig.~\ref{fig:sqw_square}(b), the peak energy along the high symmetry paths is in excellent accordance with QMC with negligible finite-size effects. 
DDMRG is also able to resolve the finite-size triple gap of $~\sim0.1$, as shown in Fig.~\ref{fig:sqw_square}(c), as well as reproducing the high-energy continuum reflecting nearly-deconfined spinon excitations~\cite{square-qmc} at $\vec q=M$ and $\vec q=S$, as shown in Fig.~\ref{fig:sqw_square}(d) and (e). 

\section{convergence of DDMRG calculation for the $J_1$-$J_2$ triangular lattice Heisenberg model}

\begin{figure}[ht]
  \includegraphics[width= 0.65\linewidth]{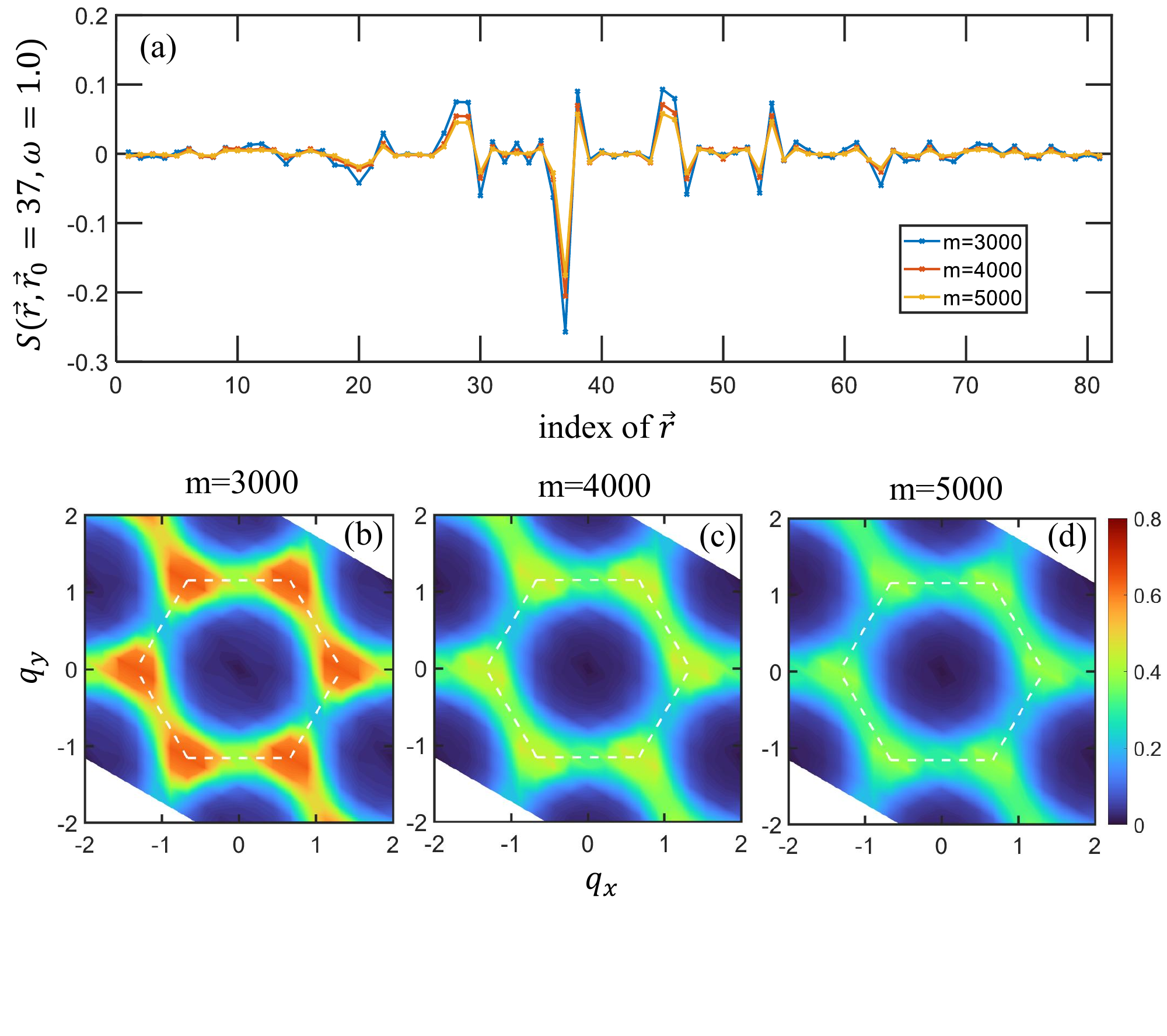}
  \vskip -1.5cm
  \caption{(a) At different bond dimension $m$ on a XC9-24 cylinder, the dynamical spin correlation function $S(\vec r,\vec r_0=37,\omega=1.0)$ of the LE state in the $J_1$-$J_2$ triangular Heisenberg model at $J_2=0.12$, using $\eta=0.1$. $\vec r$ iterates over the central $9\times 9$ region and its index equals $r_y+9(r_x-8)$. The corresponding dynamical spin structural factor $S(\vec q,\omega=1.0)$ at (b) $m=3000$, (c) $m=4000$ and (d) $m=5000$.}
  \label{fig:alg}
\end{figure}

In this section, we provide information regarding the convergence of DDMRG simulations for the $J_1$-$J_2$ triangular lattice Heisenberg model in the main text. As an example, we show a relatively challenging case of XC9-24 cylinder with $J_2=0.12$, $\omega=1.0$. The ground state has a bond dimension of 2400 (truncation error $\sim O(10^{-5})$) The dynamical spin correlation function $S(\vec r,\omega)$ (Eq.~\ref{eq:sromega}) is shown in Fig.~\ref{fig:alg}(a). While all the qualitative features are captured at $m=3000$, it converges quantitatively at $m=4000$. This is also reflected in the corresponding $S(\vec q,\omega=1.0)$, where its overall structure is formed at $m=3000$, and its peak intensity converges at $m=4000$. 

We find that this convergence behavior is generic across our simulations, that the DSSF converges when the bond dimension of the state-averaged MPS of $|Y(\omega)\rangle$, $|X(\omega)\rangle$ and $S|0\rangle$ is within twice that of the ground state (whose truncation error $\sim O(10^{-5})$), even though its typical truncation error is an order of magnitude bigger$~\sim O(10^{-4})$. 

\section{Lower-energy (LE) and higher-energy (HE) states on XC7 cylinders}

\begin{figure}[htb]
  \vskip -0.5cm
  \includegraphics[width= 0.7\linewidth]{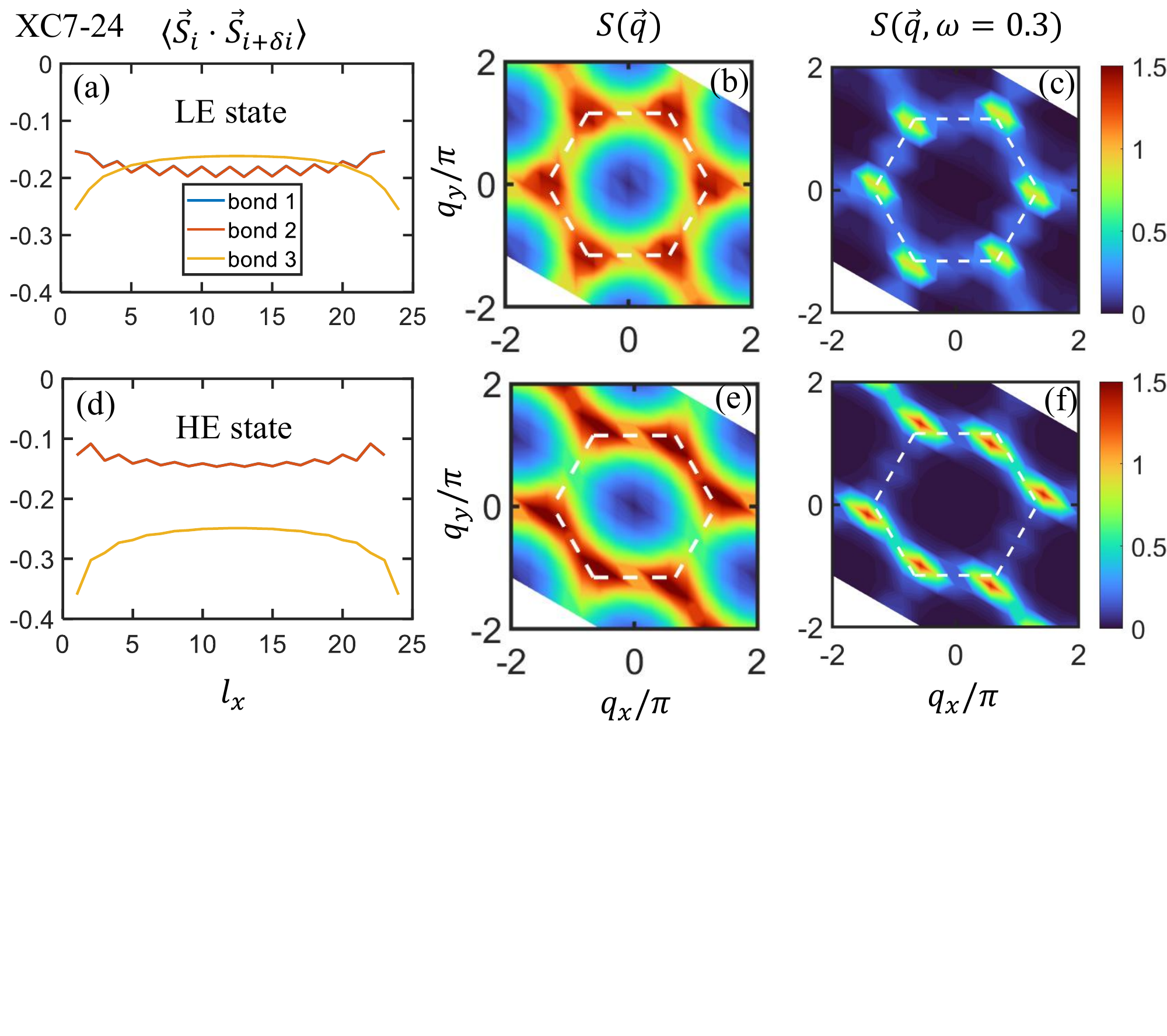}
  \vskip -3.5cm
  \caption{For the LE state (top row) and HE state (bottom row) on a XC7-24 cylinder: (a)(d) Strength of the spin-spin correlation $\langle \vec S_i\cdot \vec S_{i+\delta i}\rangle$ on three types of NN bonds (defined in Fig.1(a) in the main text), averaged over the circumferential direction and plotted versus the length of the cylinder $l_x$. (b)(e) The equal-time spin structural factor $S(\vec q)$. (c)(f) The low-energy DSSF $S(\vec q,\omega=0.3)$.}
  \label{fig:xc7}
\end{figure}

The results for LE and HE states in XC7 cylinders are shown in Fig.~\ref{fig:xc7}. Overall, the features for the LE and HE states are similar to those on XC9 cylinders in the main text: the LE state is similar to the $J_2=0$ in both the equal-time and dynamical spin structural factor, and has a substantial columnar dimer order; the HE state is different from the LE state in NN spin-spin correlation and has a more diffuse spin structural from $K$ to $M$.
Here in XC7 cylinders, we find that removing one spin on the edge for the LE state can reduce boundary effects. The ground state for both states is at $m=2100$, where the HE state is still metastable. The DDMRG calculations are at $m=2800$.

\section{LE and HE states on YC6 cylinders.}

\begin{figure}[htb]
  \includegraphics[width= 0.7\linewidth]{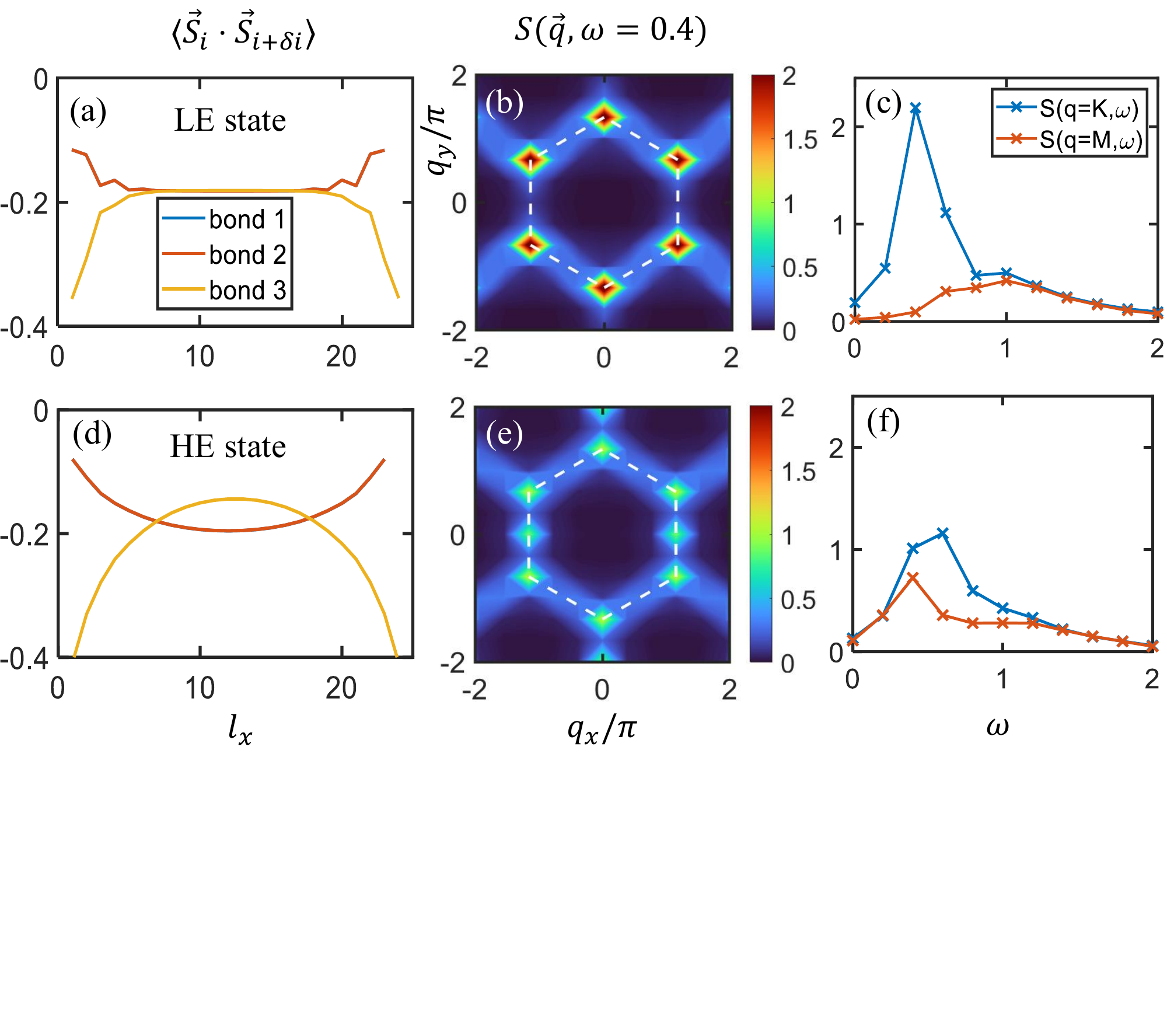}
  \vskip -3.2cm
  \caption{For the LE state (top row) and the HE state (lower row) on a YC6-24 cylinder: (a)(d) Strength of the spin-spin correlation $\langle \vec S_i\cdot \vec S_{i+\delta i}\rangle$ on three types of NN bonds (defined in Fig.1(a) in the main text), averaged over the circumferential direction and plotted versus the length of the cylinder $l_x$. (b)(e) The low-energy DSSF $S(\vec q,\omega=0.4)$. (c)(f) The DSSF at $K$ and $M$ after $D_6$ symmetrization.}
  \label{fig:yc}
\end{figure}

In this section, we show the ground states and dynamical properties of the LE and HE states on YC6 cylinders. Following the naming convention in Ref.~\cite{j1j2-zhu}, the YC cylinder has one of its NN bonds along the circumference of the cylinder. The goal of checking YC cylinders is to rule out potential finite-size effects due to the orientation of the cylinder (XC versus YC), which is supposed to be more significant on narrower cylinders. As shown in Fig.~\ref{fig:yc}, all the features of the LE and HE states on YC cylinders are identical to those on XC cylinders.

\section{Finite-size scaling of magnetic order and spin triplet gap.}
\begin{figure}[h]
\includegraphics[width= 0.6\linewidth]{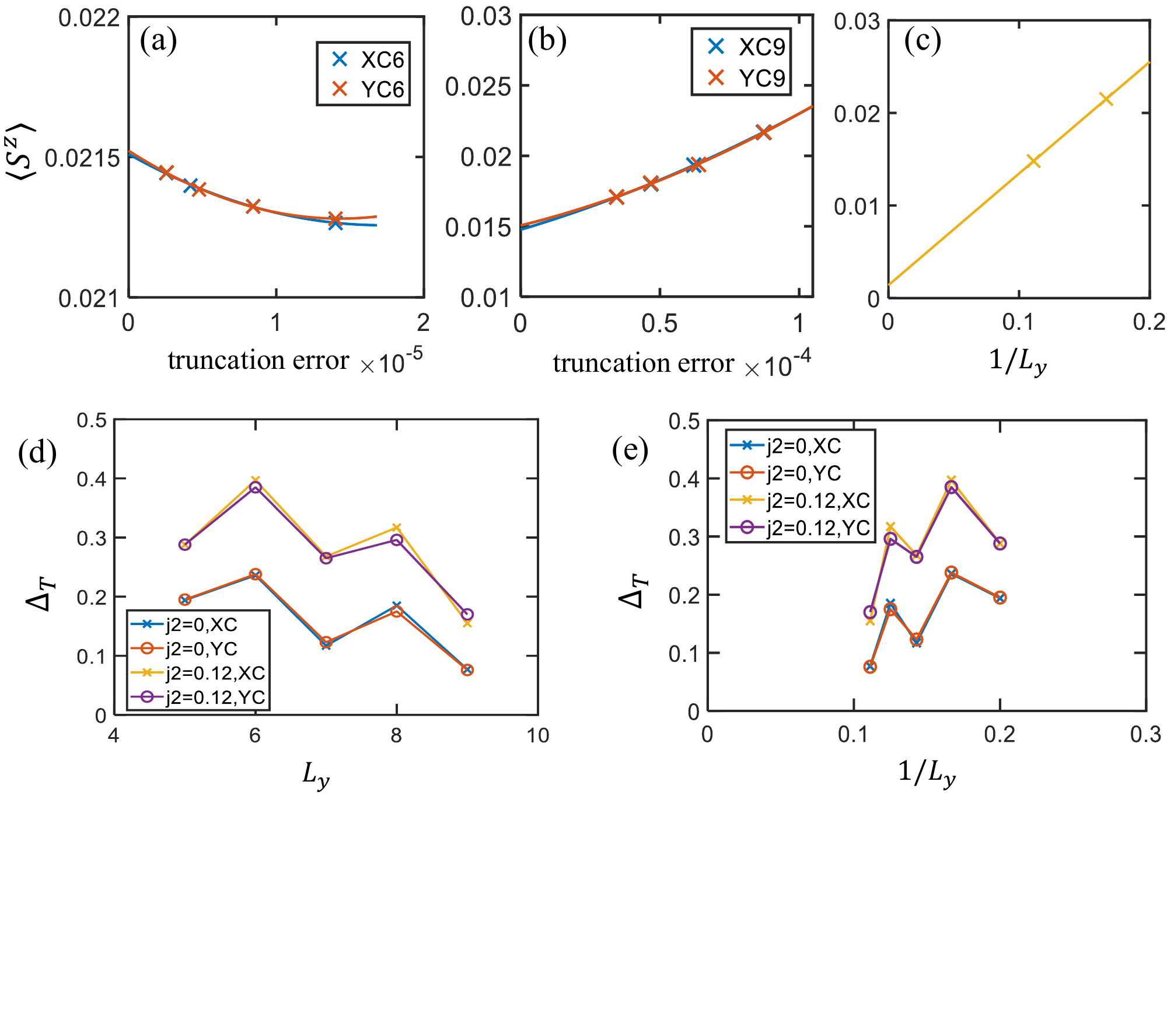}
\vskip -2.5cm
\caption{Finite-size scaling of magnetic moment for the LE state at $J_2=0.12$. (a) With edge magnetic fields pinning the 120$^\circ$ order, extrapolation of the magnetic moment $\langle S^z \rangle$ in the center of $6\times12$ XC and YC cylinders using second-order polynomials. (b) Same with (a) but on $9\times18$ cylinders. (c) $1/L_y$ scaling of the magnetic moment after averaging the extrapolated data for XC and YC cylinders in (a) and (b). (d) The spin triplet gap $\Delta_T$ for the $J_2=0$ state and the LE state at $J_2=0.12$ on $L_y$-24 cylinders, for $L_y=5-9$. (e) Same as (d) but plotted versus $1/L_y$.}
\label{fig:scaling}
\end{figure}

In this section, we first show finite-size scaling of the ordered moment for the LE state at $J_2=0.12$ using the method introduced in Ref.~\cite{steve-sasha-07}.
We use $L_y=6$ and $L_y=9$ cylinders where the 120$^\circ$ order is unfrustrated, and set their aspect ratio to be two.
To pin the 120$^\circ$ order under $U(1)$ symmetry, we apply pinning fields along the $z$ direction on the edges with pattern (+0.5,-0.25,-0.25,...), and keep the pinning fields on the two edges in phase.
The averaged ordered moment $\langle |S^z|\rangle$ in the center of the cylinder is extrapolated with truncation error using a second-order polynomial, as shown in Fig.~\ref{fig:scaling}(a) and (b) for $L_y=6$ and $L_y=9$, respectively.
We then extrapolate the obtained $\langle |S^z|\rangle$ with $1/L_y$, as shown in Fig.~\ref{fig:scaling}(c). The extrapolated zero has been used as a criterion for a magnetically ordered state~\cite{tri-sasha2,j1j2-cesar}.
Given that the extrapolated $\langle |S^z|\rangle$ is very close to zero, the magnetic order is either absent or, if present, very weak.

We also calculate the spin triplet gap $\Delta_T$ on $L_y$-24 cylinder, with $L_y$ ranging from 5 to 9, as shown in Fig.~\ref{fig:scaling}(d) and (e). For the $J_2=0$ state, overall $\Delta_T$ decreases as $L_y$ increases, with $\Delta_T$ being smaller on odd-width cylinder. This behavior is consistent with an ordered state on cylinders, which has also been observed in the coupled Heisenberg ladder~\cite{coupled-ladder-white}. For the LE state at $J_2=0.12$, we observe a similar $L_y$ dependence for the $\Delta_T$, but with a constant shift~\footnote{For the XC7 cylinder, the spin flip is attracted to the edge, so we limit the DMRG to sweep through only the central half of the system. Spin flips in other cylinders reside in the center and reflects the bulk triplet gap.}. 
This similarity suggests the LE state is probably either a weakly magnetically ordered state or a gapped $\mathbb{Z}_2$ QSL that is proximate to a continuous transition to such an ordered state.

\section {Dynamic spin structural factor under different $\eta$}
\begin{figure}[hbt]
  \includegraphics[width= 0.6\linewidth]{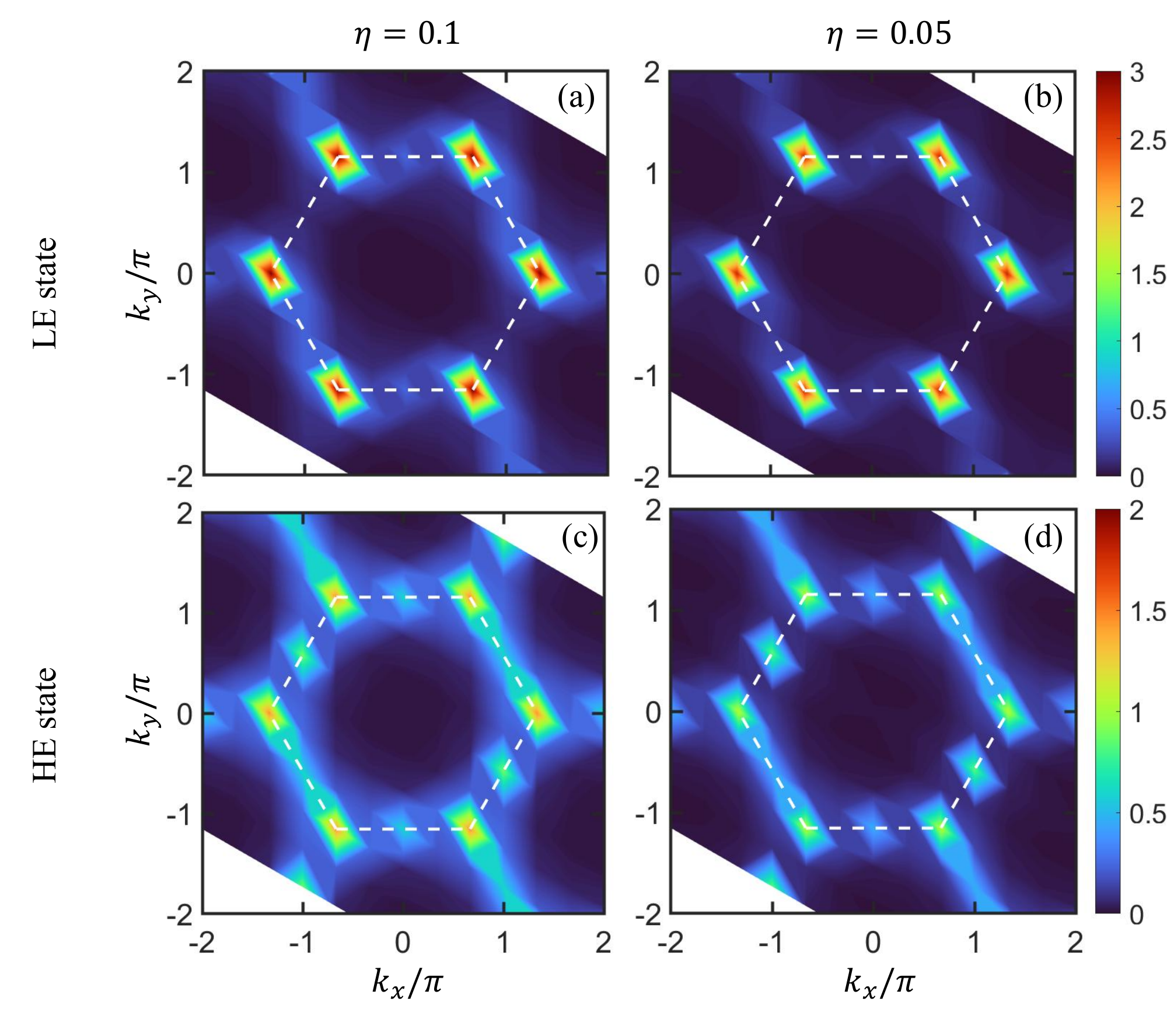}
  \vskip -0.5cm
  \caption{The DSSF at $\omega=0.4$ in XC6-24 cylinder, for the LE state (a)(b) and the HE state (c)(d), using an $\eta=0.1$ (a)(c) and $\eta=0.05$ (b)(d).}
  \label{fig:eta}
\end{figure}

In the main text of the paper, we choose a broadening factor $\eta=0.1$ in DDMRG simulations.  
In Fig.~\ref{fig:eta}, we compare the DSSF with those obtained from a smaller $\eta=0.05$. As one can see, the change of $\eta$ only induces minor quantitative change, while all the qualitative features of the LE and HE states remain the same.

\section{Symmetrized dynamical spin structural factor at low energy}
\begin{figure}[hbt]
  \includegraphics[width= 0.8\linewidth]{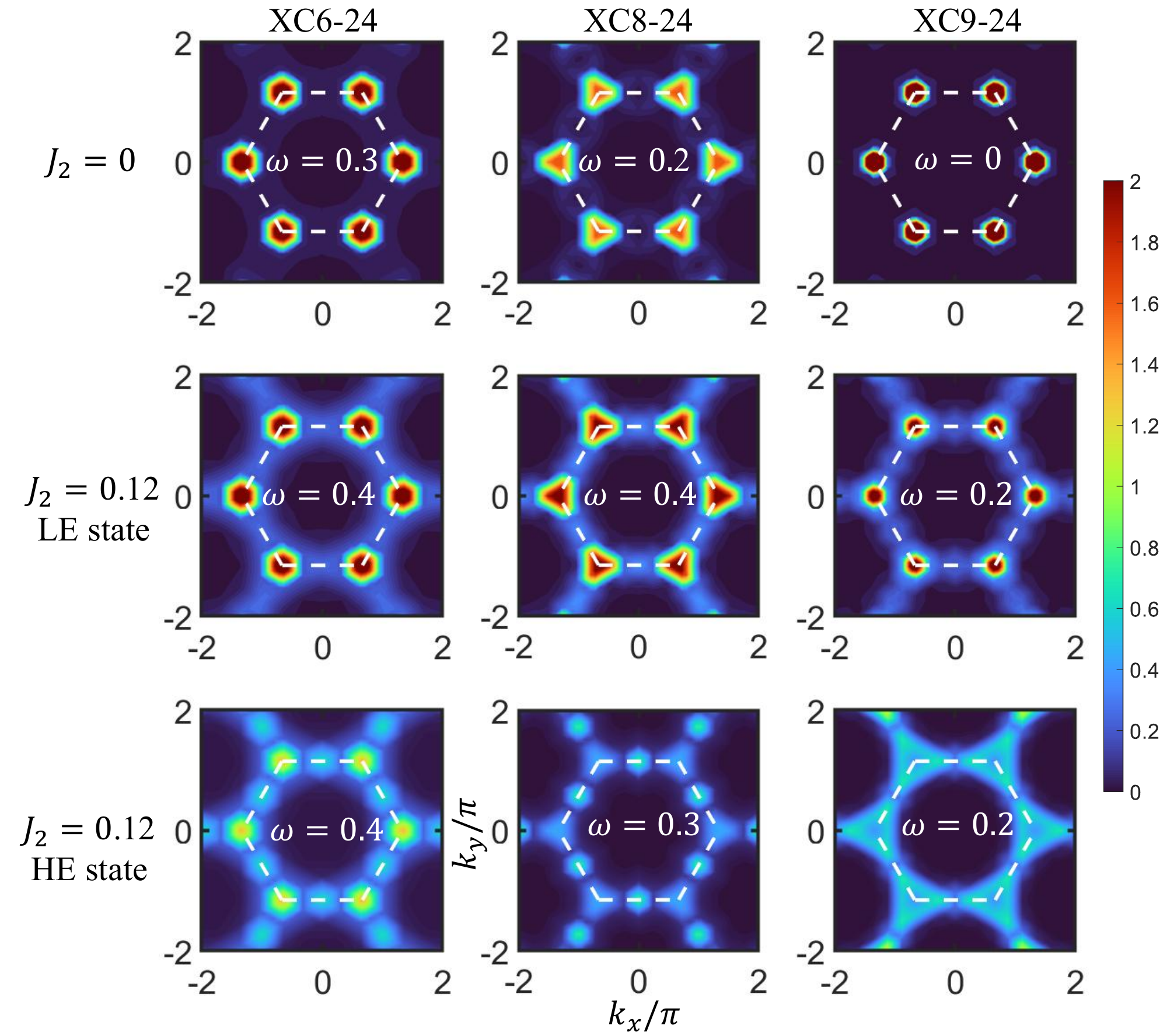}
  \caption{Low-energy DSSF (Fig.2 in the main text) after $D_6$ symmetrization.}
  \label{fig:sqw_sym}
\end{figure}

In Fig.~\ref{fig:sqw_sym}, we provide a $D_6$ symmetrized and extrapolated version of the low-energy DSSF (Fig.2 in the main text). This highlights the similarity between the $J_2=0$ state and the LE state at $J_2=0.12$ that they both have a dominant peak only at $K$, as well as the qualitative difference of the HE state at $J_2=0.12$ where the DSSF has comparable intensity from $K$ to $M$.

\bibliography{ref}